\documentclass[a4paper,12pt]{elsarticle}

\usepackage{lineno,hyperref}
\usepackage{amsmath}
\usepackage{color,soul} 
\usepackage{amssymb}    
\usepackage{graphics}
\usepackage{braket}

\modulolinenumbers[5]

\begin{document}

\begin{frontmatter}

\title{Double Chooz and a History of Reactor $\theta_{13}$ Experiments}
\tnotetext[mytitlenote]{To be printed in Nucler Physics B, Special Edition (2016).\\
~~~~~~DOI: 10.1016/j.nuclphysb.2016.04.008}

\author{Fumihiko Suekane}
\address{Research Center for Neutrino Science, Tohoku University, Sendai, 980-8578, Japan\\
Faculty of Science, Toho University, Chiba, 274-8510, Japan}
\author{Thiago Junqueira de Castro Bezerra}
\address{SUBATECH, CNRS/IN2P3, Universit\'e de Nantes, \'Ecole des Mines de Nantes, F-44307, Nantes, France}
\author{For the Double Chooz Collaboration}





\begin{abstract}
This is a contribution paper from the Double Chooz (DC) experiment to the special issue of Nuclear Physics B on the topics of neutrino oscillations, celebrating the recent Nobel prize to Profs. T.~Kajita and A.B.~McDonald. 

DC is a reactor neutrino experiment which measures the last neutrino mixing angle $\theta_{13}$. 
The DC group presented an indication of disappearance of the reactor neutrinos at a baseline of $\sim$1~km for the first time in 2011 and is improving the measurement of $\theta_{13}$. 
DC is a pioneering experiment of this research field.
In accordance with the nature of this special issue, physics and history of the 
reactor-$\theta_{13}$ experiments, as well as the Double Chooz experiment and its neutrino oscillation analyses, are reviewed.
\end{abstract}


\end{frontmatter}



\section{Introduction}
\label{sec:Introduction}
It is exciting that Profs. T.~Kajita and A.B.~McDonald are awarded the Nobel prize in physics for the discovery that neutrinos have finite mass, through neutrino oscillations.
It is an evidence that the studies of the neutrino oscillation are  appreciated for having considerably deepen our understanding of the nature. 

In short, Prof. Kajita's study was detecting atmospheric neutrino 
oscillations~\cite{98SK} and measurement of neutrino oscillation parameters, $\theta_{23}$ and $\Delta m_{32}^2$. 
Prof. McDonald's study was identifying the transformation of the flavors of solar neutrinos~\cite{02SNO} and measurement of $\theta_{12}$ and $\Delta m_{21}^2$. 
The Double Chooz (DC) experiment detects another type of neutrino oscillation using reactor neutrinos at a $\sim 1$~km baseline and is measuring a neutrino mixing parameter, $\theta_{13}$. 
This article is to explain the Double Chooz experiment as a part of the special issue of the Nuclear Physics B, for celebrating the Nobel prize. 
The physics and history of the $\theta_{13}$ measurement, an overview of the Double Chooz experiment and its results on the neutrino oscillation measurements so far are summarized in the following sections. 

\section{Neutrino Oscillation and $\theta_{13}$}
\label{sec:PhysicsTh13}
Neutrino oscillation is a phenomenon that a certain neutrino flavor periodically transforms to other flavor state. 
For the two flavor case, if there is a transition between $\nu_e$ and $\nu_\mu$, just like the transition between flavor eigenstate quarks, $d'$ and $s'$, the state equation of the neutrinos can be expressed effectively as~\cite{15Suekane}
\begin{equation}
i\frac{d}{dt}
 \begin{pmatrix}
  \nu_e \\ \nu_\mu
 \end{pmatrix}
=\frac{1}{\gamma}
\begin{pmatrix}
 \mu_e & \tau_{e\mu}^* \\
 \tau_{e\mu} & \mu_\mu
\end{pmatrix}
\begin{pmatrix}
 \nu_e \\ \nu_\mu
\end{pmatrix}, 
\label{idnu/dt=Tnu}
\end{equation}
where $\tau_{e\mu}=|\tau_{e\mu}|e^{i\phi}$ is the amplitude of the $\nu_e \leftrightarrow \nu_\mu$ cross-transition and $\mu_e$ and $\mu_\mu$ are the amplitudes of the self-transition to the original states $\nu_e \leftrightarrow \nu_e$, $\nu_\mu \leftrightarrow \nu_\mu$. 
In other words, $\mu_e$ and $\mu_\mu$ are the original masses of $\nu_e$ and $\nu_\mu$ in case $\tau_{e\mu}=0$.
$\gamma$ is the Lorentz factor which represents the time dilation of the ultra relativistically moving neutrino system. 
As a result of Eq.~(\ref{idnu/dt=Tnu}), the mass eigenstate of the neutrino, $\nu_1$ and $\nu_2$, becomes a mixture of $\nu_e$ and $\nu_\mu$: 
\begin{equation}
 \begin{pmatrix}
  \ket{\nu_1} \\ \ket{\nu_2}
 \end{pmatrix}
=
\begin{pmatrix}
 \cos\theta & -e^{-i\phi}\sin\theta \\
 e^{i\phi}\sin\theta & \cos\theta
\end{pmatrix}
\begin{pmatrix}
 \ket{\nu_e} \\ \ket{\nu_\mu}
\end{pmatrix},
\label{nu_i=MXnu_a}
\end{equation}
where $\theta$ is called the mixing angle.
The relations between the mixing angle, neutrino masses $m_1$, $m_2$ and the transition amplitudes are
\begin{equation}
 \tan2\theta = \frac{2|\tau_{e\mu}|}{\mu_\mu - \mu_e} ~~~ {\rm and} ~~~
 \begin{pmatrix}
  \mu_e \\ \mu_\mu
 \end{pmatrix}
 =
 \begin{pmatrix}
  \cos^2\theta & \sin^2\theta \\
  \sin^2\theta & \cos^2\theta
 \end{pmatrix}
 \begin{pmatrix}
  m_1 \\ m_2
 \end{pmatrix}.
 \label{Th_mua=tau_mi}
\end{equation}
The mixing angle is a measure of the ratio of the cross-transition amplitude and the difference of the original masses. 
$\mu_e$ and $\mu_\mu$ correspond to the observable mass\footnote{If we measure the mass of the $\nu_e$ state, the expectation value is $m_1 \cos^2\theta + m_2 \sin^2\theta$.} of the $\nu_e$ and $\nu_\mu$ states, respectively, 
which means that $\mu_e$ has been measured to be smaller than 
$\sim$2~eV by direct neutrino mass measurements~\cite{NeMass}.

The oscillation probability of $\nu_\mu \to \nu_e$ appearance at baseline $L$ can be calculated as 
\begin{equation}
 P_{\nu_\mu \to \nu_e}(L) = \sin^22\theta \sin^2 \frac{m_2^2-m_1^2}{4E_\nu}L,
 \label{eq: P(L)=sin22THsin2Dm2L/4E}
\end{equation}
using Eq.~(\ref{idnu/dt=Tnu}) with $\gamma = E_\nu/m_\nu$, where $m_\nu =(m_1+m_2)/2$ is the average neutrino mass.
Eq.~(\ref{eq: P(L)=sin22THsin2Dm2L/4E}) is the 2-flavor neutrino oscillation formula often used.

For the quark case, the flavor transition is caused by the Yukawa coupling to the Higgs field. 
However, the transition amplitudes for neutrinos are extremely smaller than those of quarks and what causes these transitions is not known yet. 
The standard model has to be expanded to include this phenomenon.
New physics may show up through these transitions. 
For example, they may transform neutrino to antineutrino, resulting in the Majorana neutrino state, 
or may transform known neutrinos to 4th neutrino, which is called sterile neutrino, etc.

For three flavor neutrinos,  the mixing is expressed by the Maki-Nakagawa-Sakata-Pontecorvo (MNSP) matrix, which can be parametrized by three mixing angles $\theta_{12}$, 
$\theta_{23}$, $\theta_{13}$ and one imaginary phase $\delta_{\rm CP}$:
\begin{equation}
 U_{\rm MNSP}=
 \begin{pmatrix}
  1 & 0 & 0 \\
  0 & c_{23} & s_{23} \\
  0 & -s_{23} & c_{23} \\
 \end{pmatrix}
  \begin{pmatrix}
  c_{13} & 0 & s_{13} e^{-i\delta_{\rm CP}} \\
  0      & 1  &0 \\
  -s_{13} e^{i\delta_{\rm CP}} & 0 & c_{13} 
 \end{pmatrix}
  \begin{pmatrix}
  c_{12} & s_{12} & 0 \\
  -s_{12} & c_{12} & 0 \\
  0  &  0  &  1
  \end{pmatrix},
  \label{eq:MNSP}
\end{equation}
where $s_{ij}=\sin\theta_{ij}$ and $c_{ij}=\cos\theta_{ij}$.
By fitting the results of various neutrino oscillation experiments, 
the following neutrino oscillation parameters have been calculated for the normal mass hierarchy~\footnote{If $m_1<m_3$,
it is called {\it normal mass hierarchy} and if $m_1>m_3$, {\it inverted mass hierarchy}. } case~\cite{14Capozzi}.
\begin{equation}
 \begin{split}
 \sin^22\theta_{12} &\sim 0.85,~~~\sin^22\theta_{23} \sim 0.98,~~~\sin^22\theta_{13} \sim 0.09,~~~\delta_{\rm CP} \sim 1.4\pi \\
 \Delta m_{21}^2 &\sim 7.5 \times 10^{-5}{\rm eV^2}, ~~|\Delta m_{32}^2| \sim 
 |\Delta m_{31}^2| \sim 2.5\times 10^{-3} {\rm eV^2},
 \end{split}
 \label{eq:OscillationParameters}
\end{equation}
where $\Delta m_{ij}^2 = m_i^2-m_j^2$.
%
Although the absolute neutrino mass is not known, 
if we assume $m_3 > m_2 > m_1\sim 0$, it is possible to calculate the transition amplitudes as~\cite{15Suekane}, 
\begin{equation}
 \begin{pmatrix}
  \mu_e & \tau_{e \mu}^* & \tau_{e \tau}^* \\
  \tau_{e \mu} & \mu_\mu & \tau_{\mu \tau}^* \\
  \tau_{e\tau} & \tau_{\mu \tau} & \mu_\tau
 \end{pmatrix}
 \sim
 \begin{pmatrix}
  3.8 & 1.4+4.5i & -4.4+5.1i \\
  1.4-4.5i &25 & 21 \\
  -4.4-5.1i &21 &30
 \end{pmatrix}
 {\rm meV}.
 \label{eq:3fTAM}
\end{equation}
To understand what causes this kind of transition is a central issue of the neutrino oscillation studies. 
 
The next important targets of the neutrino oscillation experiments are to detect CP violation and measure $\delta_{\rm CP}$.
 The physical CP violation effect is proportional to the Jarlskog 
invariant~\cite{85Jarlskog}: 
\begin{equation}
 J=s_{12}c_{12}s_{23}c_{23}s_{13}c_{13}^2\sin\delta_{\rm CP}.
\end{equation}
For the quark case, although $\delta_{\rm CP}$ is large, the quark mixings are small and the Jarlskog invariant is also small, 
$J_q \sim 3 \times 10^{-5} $. 
Therefore, it is considered to be difficult to explain the matter dominance of our universe by the CP violation of the quarks. 
For the neutrino case, the Jarlskog invariant is calculated as
\begin{equation}
 J_\nu \sim 0.12 c_{13}\sin2\theta_{13}\sin\delta_{\rm CP}\sim 
 0.036  \sin\delta_{\rm CP}
\end{equation}
and if $\delta_{\rm CP}$ is large, $J_\nu$ is expected to be large enough to cause the matter dominance by transferring the CP violating effects to the baryon asymmetry~\footnote{For example, see \cite{86Fukugita}.}. 
Therefore, the measurement of CP asymmetry in the lepton sector is particularly important now.
$\delta_{\rm CP}$ in (\ref{eq:OscillationParameters}) was obtained assuming the standard three flavor oscillation scheme. 
However, the CP violation effect has to be measured model-independently, from the difference between an oscillation and its CP-inverted oscillation.
One promising way to detect the CP violation is to measure the asymmetry of the oscillation probabilities between 
$\nu_\mu \to \nu_e$ appearance and $\overline{\nu}_\mu \to \overline{\nu}_e$ appearance in 
long-baseline accelerator experiments
using the relation: 
\begin{equation}
 A_{\rm CP} \equiv \frac{P(\nu_\mu \to \nu_e) - P(\overline{\nu}_\mu \to \overline{\nu}_e)}{P(\nu_\mu \to \nu_e) + P(\overline{\nu}_\mu \to \overline{\nu}_e)}
 \sim -\frac{0.087}{\sin2\theta_{13}}\sin\delta_{\rm CP}.
\end{equation}
Therefore, the value of $\theta_{13}$ is an important parameter to discuss about the feasibility of the future experiments. 
In addition, as will be explained in the next section, there are some more complications in an actual $A_{\rm CP}$ measurement. 
The reactor based $\theta_{13}$ measurement can greatly reduce such complications. 

The Double Chooz reactor neutrino experiment started aiming to measure the last unknown mixing angle $\theta_{13}$ and solve those issues.

\section{A Brief History of Double Chooz and Reactor $\theta_{13}$ Experiments}
\label{sec:History}
Regarding the history of neutrino oscillations, the atmospheric neutrino anomaly, that was the ratio of $\nu_\mu/\nu_e$ being significantly less than the naive expectation, was already reported in 1980's by Kamiokande, IMB and Soudan groups~\footnote{A good review can be seen 
at~\cite{06Losecco}.}. 
If neutrino oscillation, $\nu_\mu \to \nu_e$, caused the small $\nu_\mu/\nu_e$ ratio, its CPT inverted oscillation $\overline{\nu}_e \to \overline{\nu}_\mu$ should have taken place at the same $L/E$.
For reactor $\overline{\nu}_e$, whose energy is a few MeV, the oscillation was expected to appear at a baseline of few km~\footnote{At that time, the measured $\Delta m_{32}^2$ value was several times larger than the current one.}.  

In 1990's, the Chooz experiment in France and Palo Verde experiment in the U.S.A. tried to measure the reactor neutrino oscillation at such baselines~\cite{ChoozPaloVerde}, where the oscillation probability at the oscillation maximum is expressed as 
\begin{equation}
 P(\overline{\nu}_e \to \overline{\nu}_e) = 1- \sin^22\theta_{13}. 
 \label{eq:P=1-sin22th13}
\end{equation} 
Since both experiments could not observe significant deficit of the reactor neutrinos, 
$\theta_{13}$ was known to be small by the end of 1990's~\cite{99ChoozPaloVerde},~\footnote{Results from experimental groups are based on the two flavor oscillation scheme, but some papers performed three flavor analysis~\cite{98Narayan}.}
\begin{equation}
 \sin^22\theta_{13}<0.1. 
\end{equation}
Meanwhile, SuperKamiokande reported an evidence of the atmospheric neutrino oscillation in 1998~\cite{98SK} and the neutrino was proven to have finite mass and mixing. 
This achievement led to the 2015 Nobel prize. 
The data also indicated that the $\nu_\mu \to \nu_e$ component of the atmospheric neutrino oscillation is small, which was consistent with the small $\theta_{13}$

Being $\theta_{13}$ small, the reactor measurement was thought to be disadvantageous since it is a measurement of the small neutrino deficit while a few \% of uncertainties are included in both the expected reactor neutrino flux and absolute measurement of the flux.
Therefore, after the Chooz and Palo Verde experiments, $\theta_{13}$ was expected to be measured by the appearance experiments at 
accelerators~\cite{01T2KProposal} using the relation
\begin{equation}
 P(\nu_\mu \to \nu_e) \sim \frac{1}{2}\sin^22\theta_{13}.
 \label{eq:P(Numu2Nue)=1/2sin22Th13}
\end{equation}
This oscillation formula bases on the assumption that the solution of the solar neutrino problem is the Small Mixing Angle solution with very small $\theta_{12}$.
However, in 2002, SNO group showed the evidence of the solar neutrino flavor transition and KamLAND group showed a large reactor neutrino deficit at an average baseline of $\sim$180~km~\cite{02SNO, 03KamLAND}. 
The SNO results also led to the 2015 Nobel prize.
These results, together with other solar neutrino experiments, concluded that the solar neutrino oscillation parameter is the Large Mixing Angle solution and showed that $\theta_{12}$ is large and $\Delta m_{21}^2$ is not so small,
\begin{equation}
 \sin^22\theta_{12}\sim 0.8,
 ~~~~\Delta m_{21}^2 \sim 8\times 10^{-5} {\rm eV^2}.
 \label{eq:Th12=Dm12=}
\end{equation} 
This observation had not been expected by many physicists and changed the prospects of the neutrino oscillation studies. 
If we use the observed oscillation parameters (\ref{eq:Th12=Dm12=}) and possible $\theta_{23}$ uncertainty, the $\nu_e$ appearance probability (\ref{eq:P(Numu2Nue)=1/2sin22Th13}) becomes
\begin{equation}
 P(\nu_\mu \to \nu_e) \sim \sin^2\theta_{23} \sin^22\theta_{13}-0.04\sin2\theta_{13}\sin\delta_{\rm CP}.
 \label{eq:PNm2NwWOrm}
\end{equation}
This formula shows an important implication that if $\theta_{13}$ is not so small, there is a possibility to measure $\delta_{\rm CP}$ in future long-baseline accelerator experiments. 
Therefore, the measurement of $\theta_{13}$ became all the more important, from the mere measurement of a basic parameter to a measurement of the future possibility of detecting the leptonic CP violation. 

 However, in real experiments, the accelerator neutrino oscillation is affected by the earth matter effect and if it is included, the $\nu_e$ appearance probability (\ref{eq:PNm2NwWOrm}) becomes more complicated: 
\begin{equation}
 P(\nu_\mu \to \nu_e) \sim 
 \frac{\sin^2\theta_{23}\sin^22\theta_{13}}{(1-\rho_m L)^2}
 -0.04 \frac{\sin2\theta_{13}}{(1-\rho_m L)}\sin\delta_{\rm CP}, 
 \label{eq:PNm2NwWTrm}
\end{equation}
where $\rho_m$ comes from the matter effect.
If the neutrino energy is set so as to perform the experiment near the $\Delta m_{32}^2$ oscillation maximum, 
 $|\rho_m|\sim 0.15[{\rm /1000km}]$.
The sign of $\rho_m$ depends on either the neutrino beam is $\nu_\mu$ or 
$\overline{\nu}_\mu$. 
Therefore, $\rho_m$ introduces a fake CP asymmetry. 
The sign of $\rho_m$ also depends on the mass hierarchy (sign of $\Delta m_{32}^2$) and the fake CP asymmetry can not be corrected without knowing the mass hierarchy.
 The $\sin^ 2\theta_{23}$ term introduces another ambiguity, called $\theta_{23}$ degeneracy problem, to the oscillation probability. 
 Therefore, it is difficult to pin down all the ambiguities and to measure 
 the $\theta_{13}$ and $\delta_{\rm CP}$ by accelerator experiments only. 
 
 Meanwhile, Mikaelyan {\it et al.}~\cite{99KR2DET} proposed the near-far detectors concept to reduce the systematic uncertainty significantly for reactor based $\theta_{13}$ measurement in 1999 and solved the problem of the disappearance measurement.  
 Minakata {\it et al.}~\cite{03Minakata} pointed out the complementarity of the reactor and accelerator based $\theta_{13}$ measurements in 2003 and motivated the significance of the reactor measurement of $\theta_{13}$. 
 Huber {\it et al.}~\cite{03Huber} made thorough study of the sensitivities of the reactor $\theta_{13}$ measurement with the near-far scheme in 2003 and evidenced the effectiveness of the synergy with accelerator experiments.
 
 For reactor neutrino case, even if the values in Eq.~(\ref{eq:Th12=Dm12=}) are used, the oscillation probability does not change from Eq.~(\ref{eq:P=1-sin22th13}),  
\begin{equation}
 P(\overline{\nu}_e \to \overline{\nu}_e) \sim 
  1-\sin^22\theta_{13} +O(10^{-3}).
\end{equation}
Therefore, it is possible to measure $\theta_{13}$ by reactor neutrino experiments without ambiguities and it is ideal to measure $\theta_{13}$ by reactor experiments and combine with the accelerator data to resolve the ambiguities that intrinsically exist in the CP asymmetry measurement.
Since this type of experiments were expected to produce very important physics result with much less cost compared with accelerator based experiments, as many as thirteen projects were once proposed. 

Double Chooz~\cite{04DCLoI} is one of the pioneering projects among them inheriting the Chooz experiment. 
In February 2003, the Double Chooz made the first presentation in an international conference, NOON03 at Kanazawa, 
and in October 2003, there was the first world-wide reactor $\theta_{13}$ workshop, Future Low Energy Neutrino Workshop, at Munich and a white paper was published~\cite{04White} based on the discussions there.  
 In that workshop, the four layers detector concept was proposed by Double Chooz and KASKA groups and methodology of the high precision reactor $\theta_{13}$ experiment was mostly established.

 In March, 2004, at the midst of the Low Energy Neutrino Workshop in Niigata, the Double Chooz project was approved by the French funding agency.
At that time there were six proposed projects world-wide, Double Chooz~\cite{06DCLoI}, Kr2Det~\cite{03KR2DET} in Russia, KASKA~\cite{06KASKALoI} in Japan, Diablo Canyon~\cite{DiabloCanyon} and Braidwood~\cite{03Shaevitz} in the U.S.A., Daya Bay~\cite{07DayaBay} in China. 
Eventually, Daya Bay and newly proposed RENO~\cite{10RENO} in Korea were funded also. 
By the year of 2006, Kr2Det, Braidwood~\cite{04USDC} and KASKA joined Double Chooz and Diablo Canyon joined Daya Bay and finally the three projects proceeded with the construction of the detectors.

At the same time, accelerator based $\theta_{13}$ project, T2K, was being prepared in Japan and 
the first neutrino beam was produced in 2009.
Although the big earthquake hit north of Japan in March 2011, the T2K group successfully reported an indication of 4.5 $\nu_\mu \to \nu_e$ appearance events in June 2011~\cite{11T2KTh13} using the data taken before the earthquake and showed 
$0.03<\sin^22\theta_{13}<0.34$(90\% CL) assuming $\delta_{\rm CP}=0$.
Soon after that MINOS presented their $\nu_e$ appearance result and showed $\theta_{13}>0$ with 89\%~C.L.~\cite{11MINOSTh13}.

The Double Chooz group reported an indication of reactor $\overline{\nu}_e$ disappearance with the far detector data in November 2011 at LowNu2011 conference in Korea and the result, $\sin^22\theta_{13} = 0.086 \pm 0.041$ was published in March 2012~\cite{12DC1st}. 
Daya Bay experiment published high precision $\theta_{13}$ result using near and far detectors, $\sin^22\theta_{13} = 0.096 \pm 0.017$, in March 2012~\cite{12DayaBay1st} and the RENO experiment followed it showing 
$\sin^22\theta_{13} = 0.113 \pm 0.023$~\cite{12RENO1st}. 
Since then, the accuracy of $\theta_{13}$ is being improved and its value is playing an important role in the neutrino oscillation studies. 

\begin{figure}[htbp]
\begin{center}
\includegraphics[width=7.0 cm]{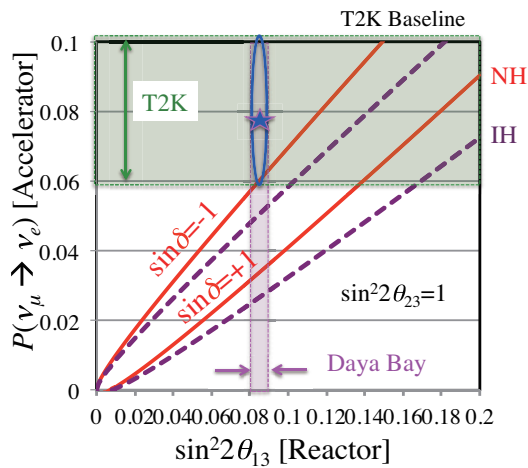}
\end{center}
\caption{\small{Synergy of the reactor and accelerator experiments. 
Horizontal axis is $\sin^22\theta_{13}$ measured by reactor experiment and vertical axis is $\nu_\mu \to \nu_e$ appearance probability measured by accelerator experiment.
The parabola regions are calculated using Eq.~(\ref{eq:PNm2NwWTrm}), assuming $\sin^2\theta_{23}=0.5$ and the T2K baseline. 
The solid parabola is for Normal Hierarchy and the dashed parabola is for Inverted Hierarchy.
The shaded areas are $1\sigma$ bands of Daya Bay~\cite{15DayaBay} and T2K~\cite{14T2K} experiments.
}}
\label{fig:R+A}
\end{figure}

\begin{figure}[htbp]
\begin{center}
\includegraphics[width=10.0 cm]{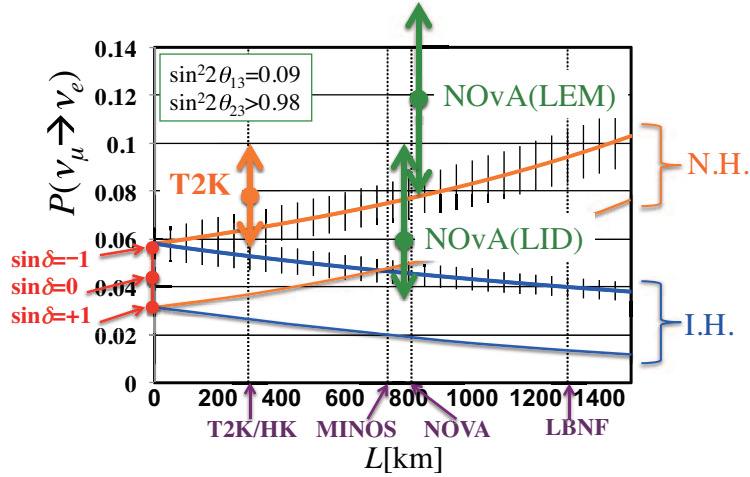}
\end{center}
\caption{\small{Baseline dependence of the $\nu_\mu \to \nu_e$ appearance probability at the $\Delta m_{32}^2$-oscillation maximum.
The horizontal axis is the baseline and the vertical axis is the $\nu_e$ appearance probability. 
The four lines correspond to $\sin\delta = \pm 1$ and mass hierarchy cases.
The thin error bars on the $\sin\delta=-1$ lines show uncertainty from the $\theta_{23}$ degeneracy. 
The data are $1\sigma$ range of T2K~\cite{14T2K} and NOvA~\cite{15NOVA} results.
NOvA showed two results from different analysis methods (LEM and LID). }}
\label{fig:T2K+NOVA}
\end{figure}

\begin{figure}[htbp]
\begin{center}
\includegraphics[width=9.0 cm]{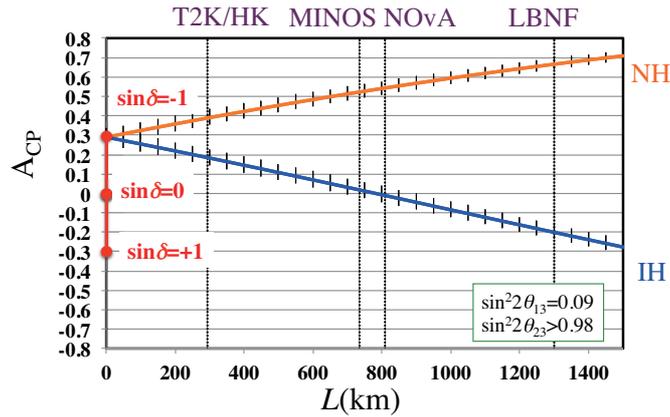}
\end{center}
\caption{\small{Baseline dependence of the CP asymmetry, $A_{\rm CP}$, at the $\Delta m_{32}^2$-oscillation maximum.}}
\label{fig:CPVLine}
\end{figure}

Fig.~\ref{fig:R+A} shows an example of the strong synergy effect of the reactor and accelerator experiments. 
The overlap of the allowed regions indicates that the normal hierarchy and $\sin\delta_{\rm CP} =-1 $ are slightly preferred. 
This kind of discussion could not be made if there were only one type of experiment.
Using the measured $\theta_{13}$, it has become possible to calculate the baseline dependence of the $\nu_\mu \to \nu_e$ appearance probability with the matter effect at the oscillation maximum ($E/L=\Delta m_{32}^2/2\pi$) as shown in Fig.~\ref{fig:T2K+NOVA}. 
The relation of T2K and NOvA results can be clearly comprehended from this figure. 
A similar calculation can be made for $A_{\rm CP}$ as shown in Fig.~\ref{fig:CPVLine}.
The data from long-baseline accelerator experiments with different baselines can be combined along a line to measure $\sin\delta_{\rm CP}$ and the mass hierarchy at once. 


\section{The Double Chooz Experiment}
\label{sec:DC}
\subsection{The Detector}
Since the $\overline{\nu}_e \to \overline{\nu}_e$ disappearance probability expected was $\sim$10~\% or less, it was necessary to design the neutrino detector in such a way that the $\overline{\nu}_e$ deficit can be measured with an accuracy of $ \sim 1 \%$ level.
Therefore, several techniques which significantly reduce the systematic uncertainties are employed in the Double Chooz detector.

\begin{figure}[htbp]
\begin{center}
\includegraphics[width=7.0 cm]{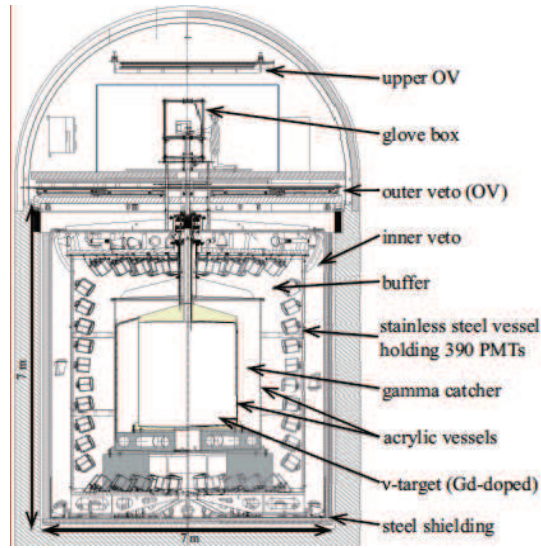}
\end{center}
\caption{\small{Double Chooz detector. 
From inner to outer, 
(i) Neutrino target scintillator,
(ii) $\gamma$-catcher scintillator,
(iii) Buffer oil,
(iv) PMT array,
(v) Inner muon veto scintillator and 
(vi) Outer muon veto counter.
}}
\label{fig:DC_detector}
\end{figure}

Double Chooz experiment uses the underground laboratory, which was used for the Chooz experiment, for the far detector. 
The rock overburden is $\sim$300~meter-water equivalent. 
Fig.~\ref{fig:DC_detector} shows a schematic view of the Double Chooz detector~\cite{DCDetector}. 
The central part (i) is the neutrino target which consists of a Gadolinium (Gd)-loaded liquid scintillator~\cite{DCLS}, with a 10~m$^3$ volume, contained in an acrylic vessel. 
The second liquid layer (ii) is Gd-free liquid scintillator nicknamed $\gamma$-catcher, which detects $\gamma$-rays escaping from the target region. 
The light output efficiency is tuned to be the same as the neutrino target in order to reconstruct the neutrino and Gd signal energy precisely. 
The volume is 23~m$^3$ and it is contained in the second acrylic vessel.
The third liquid layer (iii) is a non scintillating oil called the buffer oil, which shields the background neutrons and $\gamma$-rays from outside and PMT glass. 
The volume is 110~m$^3$ and it is contained in a stainless steel vessel. 
(iv) 390 10 inch low-background PMTs~\cite{DCPMT}, which is mounted inside the stainless steel vessel, detect the scintillation lights from the neutrino target and the $\gamma$-catcher. 
The set of the subdetectors: (i) to (iv) defines the Inner Detector (ID).
The forth liquid layer (v) is another liquid scintillator, which actively vetoes the cosmic-ray background. 
Its volume is 90~m$^3$ and contained in a stainless steel vessel. This layer is named the Inner Veto (IV) and it is optically separated from the ID~\cite{DCIV}.
There are 15~cm thick demagnetized iron shields outside the tank. 
Finally the whole upper part of the detector is covered by a large-area plastic scintillator array (vi), called Outer muon Veto (OV), which vetoes the cosmic rays that do not hit the active detector region.
The readout system~\cite{DCReadout} uses 500~MHz 8~bit flash ADC to keep wave forms for later detailed analyses.

\subsection{Neutrino Detection}
The reactor $\overline{\nu}_e$ collides with a proton in the liquid scintillator and performs the inverse beta decay (IBD) reaction,
\begin{equation}
 \overline{\nu}_e + p \to e^+ + n.
 \label{eq:ibd}
\end{equation}
The diagram of the IBD event is the inverse of the neutron decay. 
Since the $q^2$ of the reactor neutrino event is small, the recoil and radiative corrections are small and the absolute cross section can be obtained precisely from the neutron lifetime.  

The positron emits scintillation light and slows down until it annihilates with an electron existing around. 
Therefore, there is a minimum energy ($2m_e = 1.02$~MeV) for the neutrino signal.
By setting the energy threshold below this energy, the error of the detection efficiency from the threshold uncertainty can be eliminated. 
Since the recoil energy of neutron is very small, the original neutrino energy can be obtained from the energy of the positron signal, 
\begin{equation}
 E_{\overline{\nu}} \sim  E_{e^+} + 0.8~{\rm MeV}.
\end{equation}
This property enables the spectrum analysis of the oscillation. 

The produced neutron thermalizes quickly by colliding with protons around. 
Approximately 30~$\mu$s later in average, the thermal neutron is absorbed by Gd and the excited Gd isotope emits $\gamma$-rays whose total energy is 8~MeV. 
\begin{equation}
 n + Gd \to Gd'+\gamma s (\Sigma E_\gamma \sim 8~{\rm MeV})
\end{equation}
The $\gamma$-rays which escape from the target region are captured by the $\gamma$-catcher scintillator and the original energies of the neutrino and the Gd signals can be obtained by the sum of the  scintillation lights from both the neutrino target and the $\gamma$-catcher. 
By setting the energy threshold of Gd signal to be well below the 8MeV peak, the uncertainty of the detection efficiency of the Gd signal can be reduced much. 
The Gd signal happens only in the neutrino target. 
The neutrino signal is identified by the delayed coincidence of the positron signal and the Gd signal.
This means that the fiducial volume cut based on position reconstruction of the signals is not necessary.  
Therefore, the uncertainty associated with the fiducial volume definition can be avoided. 

For some analyses, we use hydrogen signals requiring 
\begin{equation}
n+H \to d + \gamma (2.2{\rm MeV})
\end{equation} 
as the delayed signals in addition to the Gd signals.
For those analyses, the systematic uncertainty is slightly worse than the Gd analyses but the statistic becomes 3 times larger. 
Since the events in the Hydrogen sample are completely different from the Gd sample events and independent, the accuracy of the $\theta_{13}$ measurement improves by combining both information. 

\subsection{The Near-Far detectors concepts}
The Double Chooz experiment measures the $\overline{\nu}_e$ from the Chooz nuclear power station. 
The Chooz power station has two new-generation pressurized water reactors (PWR) with a thermal power of 4.25~GW each.
As explained before, DC uses two detectors with the identical structure, one near the oscillation maximum (1.05~km) and the other closer to the reactors (400~m) where the oscillation is still small. 
By comparing the data from the near and the far detectors, the systematic uncertainties of reactor neutrino flux, neutrino detection efficiency and detector mass, can be largely suppressed. 
The two detectors are located close to the iso-flux line, on which the ratios of the squared distances to the two reactors are the same for the near and far detectors and the neutrino flux uncertainty caused by possible property differences between the two reactors are also canceled. 

The far detector was completed in 2010 and has been taking data since then.
It was possible to measure the background directly since both reactors simultaneously turned off two times.
The construction of the near detector was finished in 2014.
At the time of this report, the near detector has taken one year of data and the physics results with both the detectors are expected to be reported within 2016. 

\section{Physics Results}
\label{sec:Physics}

As already noted, the Double Chooz collaboration was the first one among the neutrino reactor experiments to present an indication of non-zero $\theta_{13}$ at LowNu11 conference back on November 2011. This result was published later in 2012 in Phys. Rev. Lett.~\cite{12DC1st}. The first results of Daya Bay and RENO experiments were also published in the same volume~\cite{12DayaBay1st,12RENO1st}. At that time, 96.8 days of data-taking were used, leading to 4121 IBD candidates, where 4344 $\pm$ 165 events, including backgrounds, were expected. This lack of number of events was interpreted as due to neutrino oscillations with a relatively large $\theta_{13}$ value. The no oscillation hypothesis was excluded at the 94.6\% C.L. Since then, many efforts and advances were performed by the collaboration to improve this result and in this section we summarize them until the current time, defining each DC data-release as: Gd-I~\cite{12DC1st}, Gd-II~\cite{12DCTH13}, H-II~\cite{13DCTH13}, Gd-III~\cite{14DCTH13IMP} and H-III~\cite{15DCTH13}. More details on the analysis and results can be found on these references and in those cited within the next sections.

\subsection{IBD Prediction}
\label{sec:IBDpredic}

DC, together with the current reactor neutrino experiments, uses the IBD reaction, Eq.~(\ref{eq:ibd}), the same as the classic Cowan-Reines first observation of the neutrino. To predict the number of IBD events, all the simulation chain from the nuclear fission process in the reactor core, to the interaction in the liquids, until the event read out by the electronics is performed. At the reactor level, the MURE code, fed by the \'Electricit\'e de France S.A. (EdF) parameters, such as the thermal power, computes the fissile isotope composition and evolution through time~\cite{DCReactor}. The fission rates are then used together with the measured and calculated neutrino spectra~\cite{NuFlux} and IBD cross section to draw the neutrino energy. Then, both positron and neutron energy deposition, interaction process and light emission by the scintillator are simulated with a Geant4 based code with a custom neutron scattering model. A custom made package deals with the electronic pulse, which is the output signal from the PMTs, passing by the front-end electronics and recorded by the flash-ADC boards~\cite{DCDetector,DCPMT,DCReadout}. The final MC information has the same format as the data and all the reconstruction (pulse, vertex position, and energy) algorithms are applied in the same way to both sets. A high precision in this prediction is sought for since in the first phase of the experiment, only the far detector is used. Finally, to avoid uncounted factors on the reactor neutrino flux (such as model imprecision and/or short-ranged oscillations), DC uses the Bugey4 mean cross-section per fission measurement as an anchor point~\cite{Bugey4}, where the difference of the reactor fuel compositions are taken into account.

\subsection{Backgrounds}
\label{sec:BG}
	

Knowing precisely the backgrounds involved in a neutrino detector is very important for oscillation analysis. This is especially important in Double Chooz due to relatively shallow overburden. However, several methods to predict, measure, cross-check and veto the backgrounds were developed in DC by the analysis efforts.

The backgrounds of DC are divided in three types: accidentals, cosmogenic and correlated. Accidentals are two uncorrelated signals that pass the IBD selection (see Sec.~\ref{sec:IBDselect}) mimicking a true neutrino signal. An example is a gamma radiation from the environment followed by a neutron, produced by spallation interaction from cosmic muons, captured a few micro-second later. Its rate is easily estimated by the combination of the prompt and delayed single signal rates. Moreover, it can be precisely measured by an off-time method, where after the prompt event, an offset time window of one second is applied to search for the delayed candidate (adding more windows for this offset increases the statistics of the method). In this way, not only the accidental rate, but its prompt signal energy spectrum is measured with high precision.

When a muon passes through the detector's surrounding rocks, without being tagged by either the OV or IV, it can produce multiple neutrons, by spallation process, that can enter the detector. Since it is scintillator based, scatterings of protons by a neutron can produce a prompt signal, while the neutron is thermalized and might be captured also inside the detector within the coincidence time window, characterizing a correlated background. A second process of such background is the stopping-muons, since there is an acceptance hole at the detector chimney, used for source calibration, where a muon can enter, deposits an amount of energy (prompt) and stops, decaying with an electron emission (delayed). The OV and IV are powerful tools to measure the rate and energy spectra of these correlated events. While their main function is to tag muons entering the ID in an effective way, the IV is also used to get information of neutrons that also might enter the detector.

The last component of the backgrounds, the cosmogenic one, is composed of long-lived $\beta$-n (beta-neutron) emitters. They are the isotopes $^{9}$Li, mainly, and $^{8}$He that are produced inside the ID when a high energy muon interacts with the carbon of the scintillator, or other heavy nuclei. Since they have a life-time longer than 100~milliseconds, a time veto after such muons would increase the dead-time of the detector considerably. However, fits to the time correlation between IBD candidates and the previous muon, and likelihood estimations, can be done to evaluate this background contamination.


Giving that the DC uses only two reactors as a neutrino source, there is a good chance that both reactors stop to operate for a while. This is the perfect opportunity for a background measurement and confirmation of the background models described above. In fact, on the last four years, about seven days of data taking was performed at this configuration~\cite{14DCBKG}, and the measured data agreed with the background models of all the data releases.


\subsection{IBD Candidates Selection}
\label{sec:IBDselect}

To select the candidates of an IBD in the recorded data, we look for positron and neutron capture event pairs correlated in time and space, which strongly suppress the backgrounds caused by single events. Thus, suitable selection cuts on energy, time difference and distance between reconstructed vertexes are applied for the event pair. The cuts depend on which type of analysis is being performed: neutron captures on Gadolinium isotopes (n-Gd, and main one) or on Hydrogen (n-H), where the main difference is on the delayed energy, defined by the gammas released after the capture. An extra cut is the multiplicity, which selects pairs that have only the prompt and the delayed event in a fixed time window, centred at the prompt event to reject multiple neutron captures. 

Additional variables and cuts are used to further reduce the background contamination 
(described in the previous section) 
keeping the maximum amount of signal as possible. These multiple vetoes can be summarized as: $\rm F_{V}$, OV, IV, Li+He, ANN and MPS vetoes, and they are briefly described in the next paragraphs. All the IBD selection cuts and vetoes are summarized in Table~\ref{tab:SelectionComp} for all the DC data releases, where it can also be seen when new criteria were developed.

\begin{table}[htbp]
\centering
\caption{\small{Comparison between the IBD candidate selection method for the DC data-sets released so far. The circle ($\bigcirc$) means that a veto or cut is in use, while the cross ($\times$) means it is not.}}
\resizebox{\columnwidth}{!}{%
\begin{tabular}{lccccc}
\hline
\hline
						& Gd-I	 (2011)	&	Gd-II (2012)	&  H-II	 (2013)	&	Gd-III (2014)&	H-III (2015)	\\
\hline
Prompt [MeV]				&	[0.5, 12.2]	&	[0.7, 12.2]	&	[0.7, 12.2]	&	[0.5, 20]	&	[1.0, 20]	\\
Delayed [MeV]			&	[6.0, 12.0]	&	[6.0, 12.0]	&	[1.5, 3.00]	&	[4.0, 10]	&	[1.2, 3]		\\
Time Correlation [$\mu$s]&	[2.0, 100]	&	[2.0, 100]	&	[10, 600]	&	[0.5, 150]	&	[0.5, 800]	\\
Space Correlation [cm]	&	$\times$		&	$\times$		&	$<$ 90		&	$<$ 100		&	$<$ 120		\\
Multiplicity		[$\mu$s]	&	[-100, 400]	&	[-100, 400]	&	[-600, 1000]	&	[-200, 600]	&	[-800, 900]	\\
$\rm F_{V}$ veto			&	$\times$		&	$\times$		&	$\times$		&	$\bigcirc$		&	$\bigcirc$		\\
OV veto					&	$\times$		&	$\bigcirc$		&	$\bigcirc$		&	$\bigcirc$		&	$\bigcirc$		\\
IV veto					&	$\times$		&	$\times$		&	$\times$		&	$\bigcirc$		&	$\bigcirc$		\\
Li+He veto				&	$\times$		&	$\times$		&	$\times$		&	$\bigcirc$		&	$\bigcirc$		\\
HE Muon Veto				&	$\times$		&	$\bigcirc$		&	$\times$		&	$\times$		&	$\times$		\\
MPS						&	$\times$		&	$\times$		&	$\times$		&	$\times$		&	$\bigcirc$		\\
ANN						&	$\times$		&	$\times$		&	$\times$		&	$\times$		&	$\bigcirc$		\\
%
%
\hline
\hline
\end{tabular}%
}
\label{tab:SelectionComp}
\end{table}

$\rm F_{V}$ is the likelihood output of the vertex reconstruction algorithm used by DC. It tells how likely an event is to be a point-like source. Stopping-$\mu$ events in the chimney tend to show different hit pattern than a point-like source and hence $\rm F_{V}$ becomes large for such events.	

OV veto (OVV) is based on the independent detector (OV) with a dedicated DAQ system. IBD candidates that are time coincident with an OV signal are rejected. 

IV veto (IVV) takes advantage of the fact that the IV is an active volume surrounding the ID. IV also handles good vertex and timing hit reconstruction algorithms and are examined to enhance the rejection of accidental background as explained later in this section. This information is combined with the ID to reject coincident events, since fast neutrons (fast-n) entering the detector from outside can deposit some energy in the IV before mimicking an IBD event. While in the Gd analysis only the IV information is checked for the prompt candidate, in the Hydrogen analysis both prompt and delayed IV data are examined to enhance the rejection of accidental background as explained later in this section. 

The Li+He veto is based on a likelihood calculation for each prompt event and the preceding muon, that takes into account the distance of the vertex position from the muon track, reconstructed by an especially developed algorithm~\cite{DCMuonReco}, and the number of neutron candidates following the muon within one millisecond. Prompt signals that satisfy the cut condition are rejected as cosmogenic background, where the maximum likelihood is chosen from all combinations with the preceding muon within 700 milliseconds.


Multiplicity Pulse Shape (MPS) veto was designed to tag fast-n events as well. They rely on the FADCs recorded pulses to identify small energy deposits in the ID, which can be due to proton recoils before the main signal in the FADC window. 

Recently, the DC collaboration released a new result on the Hydrogen neutron capture analysis~\cite{15DCTH13}, where the accidental is the main background but now highly suppressed when compared with the previous analysis and other experiments. This suppression comes from two new improvements: the first one is a multi-variable analysis based on Artificial Neural Networks (ANN), which takes advantage on the fact that the distributions of the delayed energy and time and space correlation between prompt and delayed of the accidental events differ from the IBD ones. The second achievement is the use of the IV, as explained above, to tag environmental high energy gammas (mainly from the 206 Thallium isotope) coming from the rocks that contaminates the prompt and delayed signals in case of n-H candidates. This tagging method showed to be effective by reducing 25\% of the accidentals after the ANN is already applied.

\begin{figure}[htbp]
\begin{center}
\includegraphics[width=0.49\textwidth]{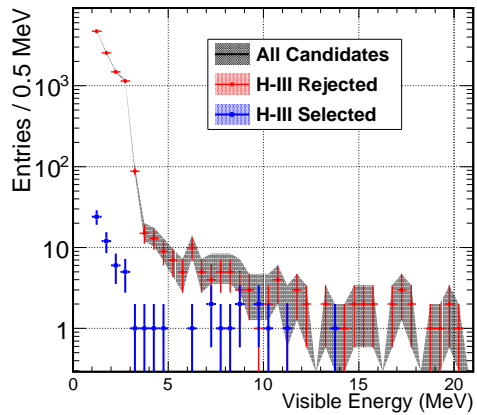}
\end{center}
\caption{\small{Reactor off spectrum for the n-H analysis, where red points are the reject events by the vetoes and the blue points are the background events that passed all the selection criteria. A two order of magnitude of rejection power can be observed.}}
\label{fig:offoff2}
\end{figure}

The power of the vetoes can be seen on the plot of Fig.~\ref{fig:offoff2}, where they are applied to the reactor-off data, of the n-H analysis, reducing the background amount by two orders of magnitude. 

\begin{figure}[htbp]
\begin{center}
\includegraphics[width=0.45\textwidth]{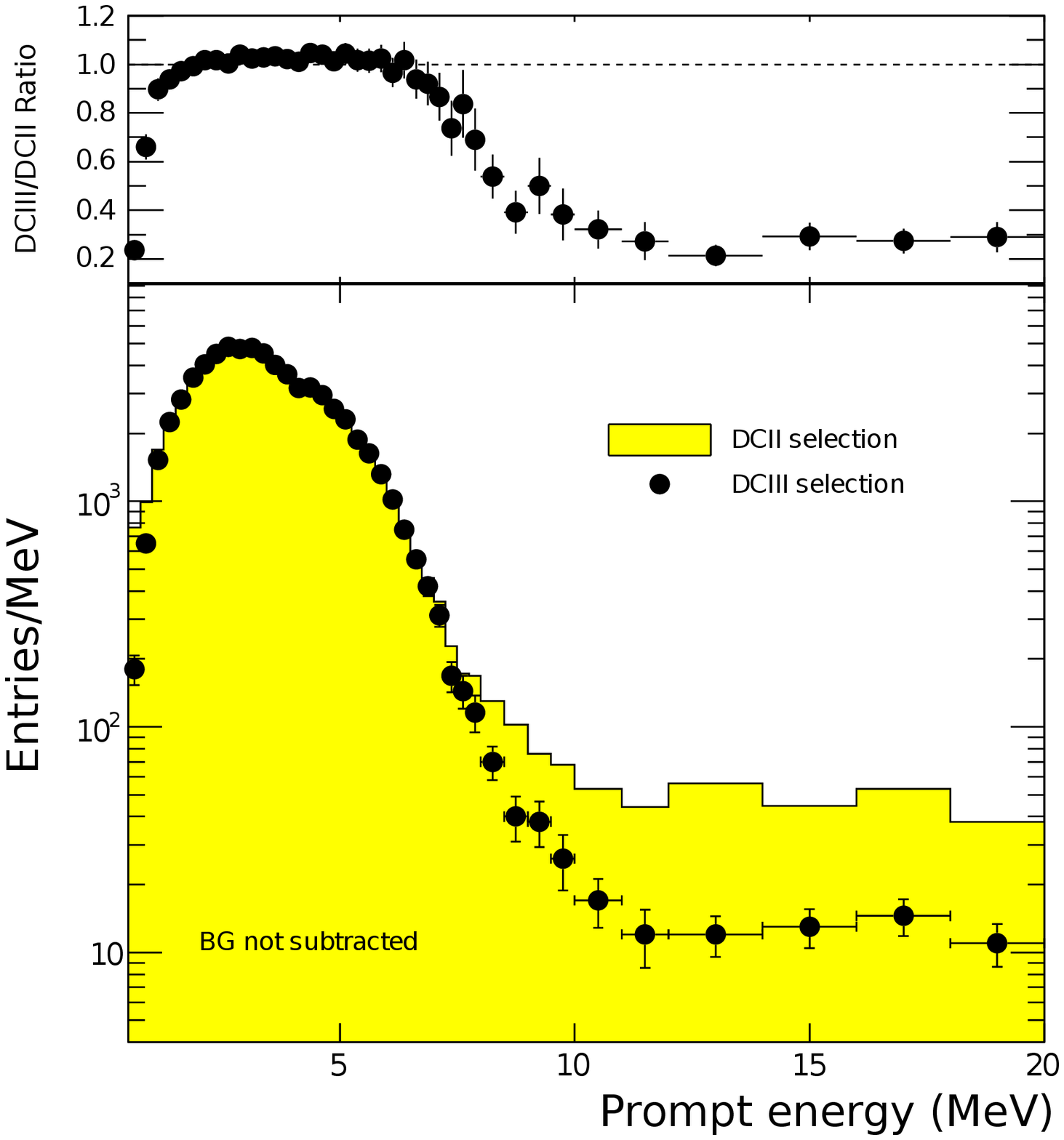}
\includegraphics[width=0.45\textwidth]{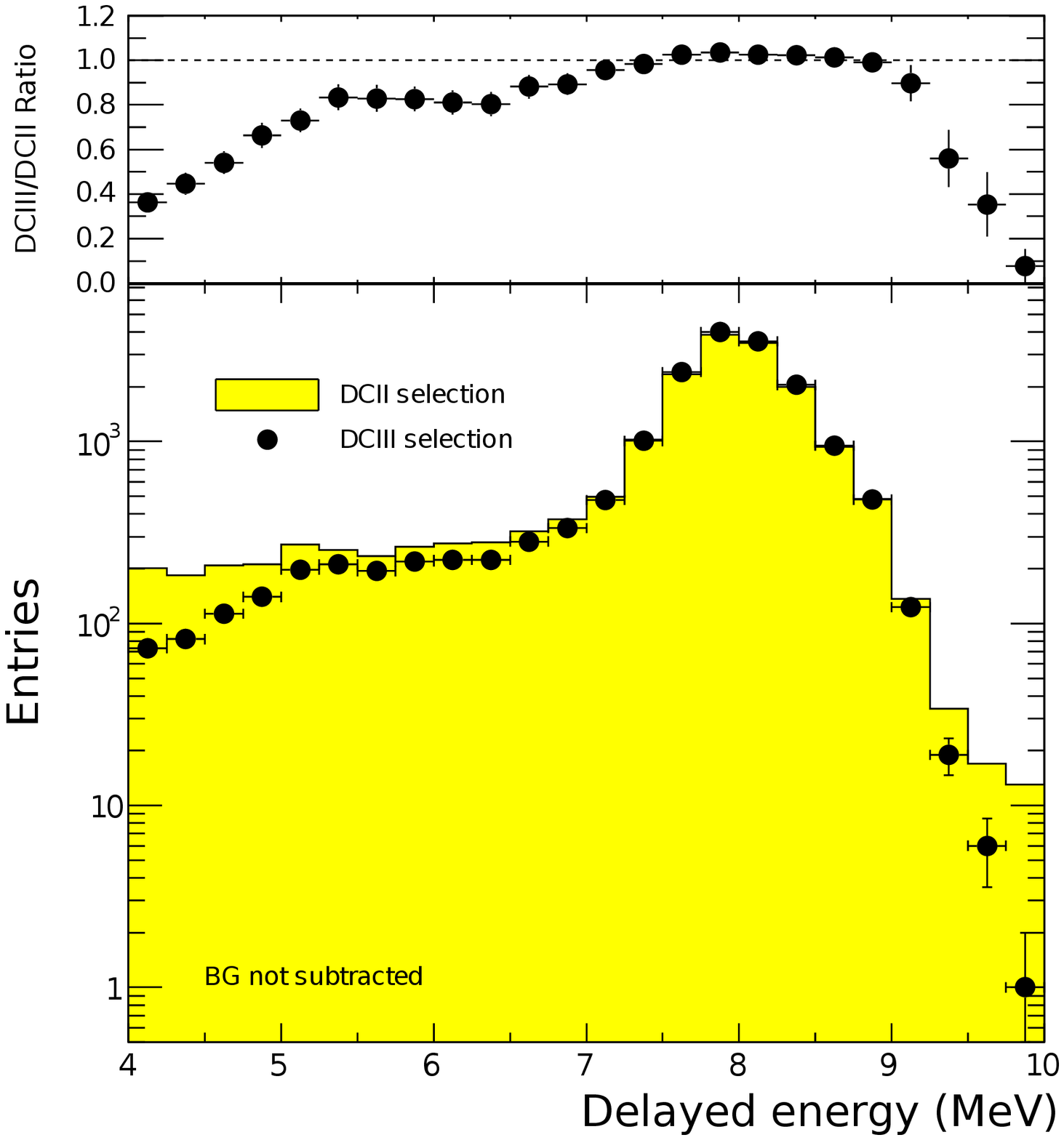}\\
\includegraphics[width=0.45\textwidth]{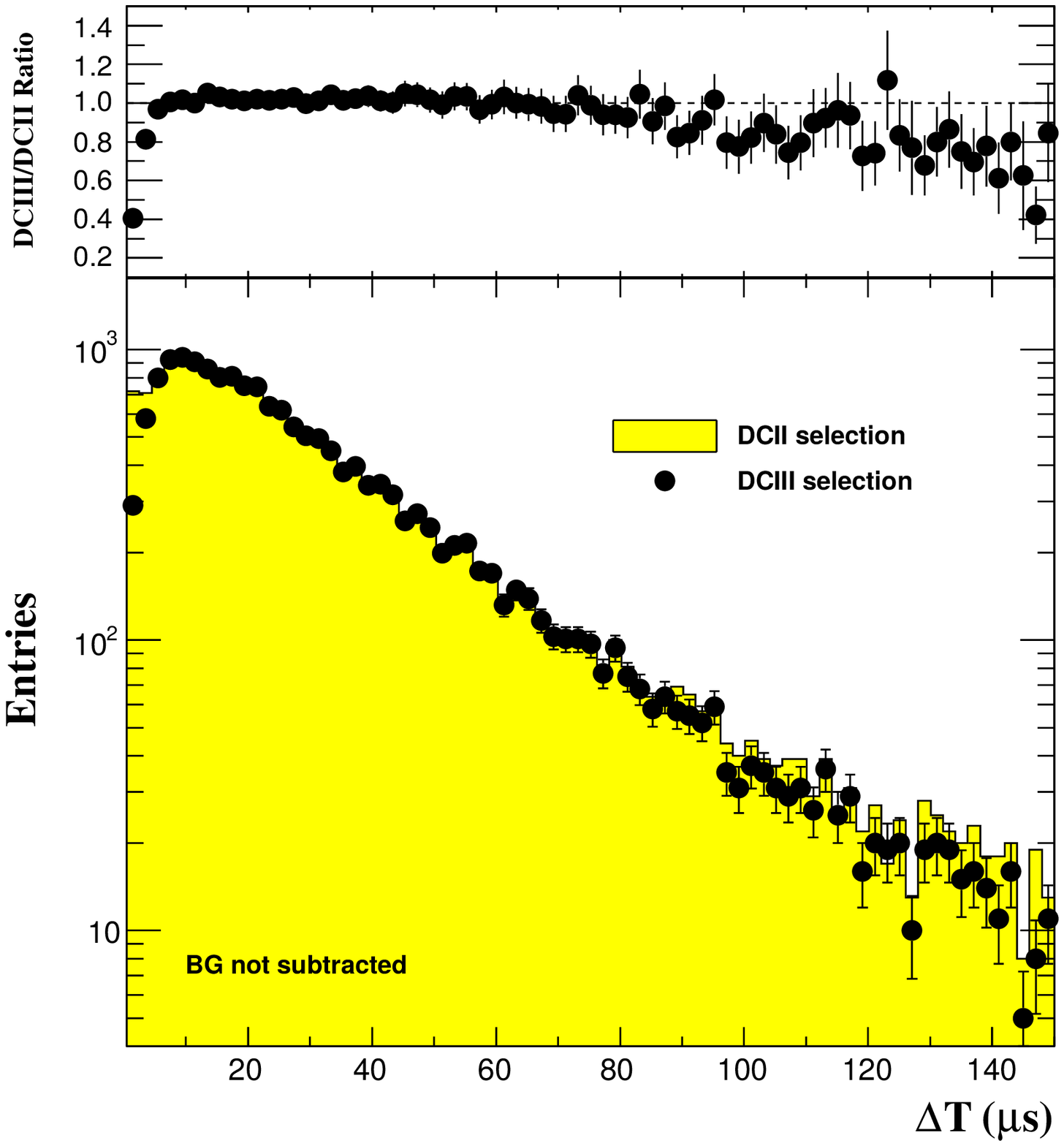}
\end{center}
\caption{\small{Comparison between the previous (Gd-II, yellow histograms) and actual (Gd-III, black circles) IBD selection methods, for the prompt (upper-left) and delayed (upper-right) energy spectra, and the time correlation (bottom center) distributions. The insets above each plot show the Gd-III/Gd-II ratio. The improvement can be seen in the critical region of the plots (low and high prompt energy and low time correlation), where the new analysis selects less events, keeping the signal part almost unchanged.}}
\label{fig:SelectionComp}
\end{figure}

Besides the new methods developed, H-III also took advantage of the achievements of the Gd-III analysis methods, as seen in Table~\ref{tab:SelectionComp} when compared with H-II. The table also shows that while all the cut windows were enlarged, many vetoes were also included, which improved the IBD detection efficiency, as seen in Fig.~\ref{fig:SelectionComp}.

\begin{figure}[htbp]
\begin{center}
\includegraphics[width=0.49\textwidth]{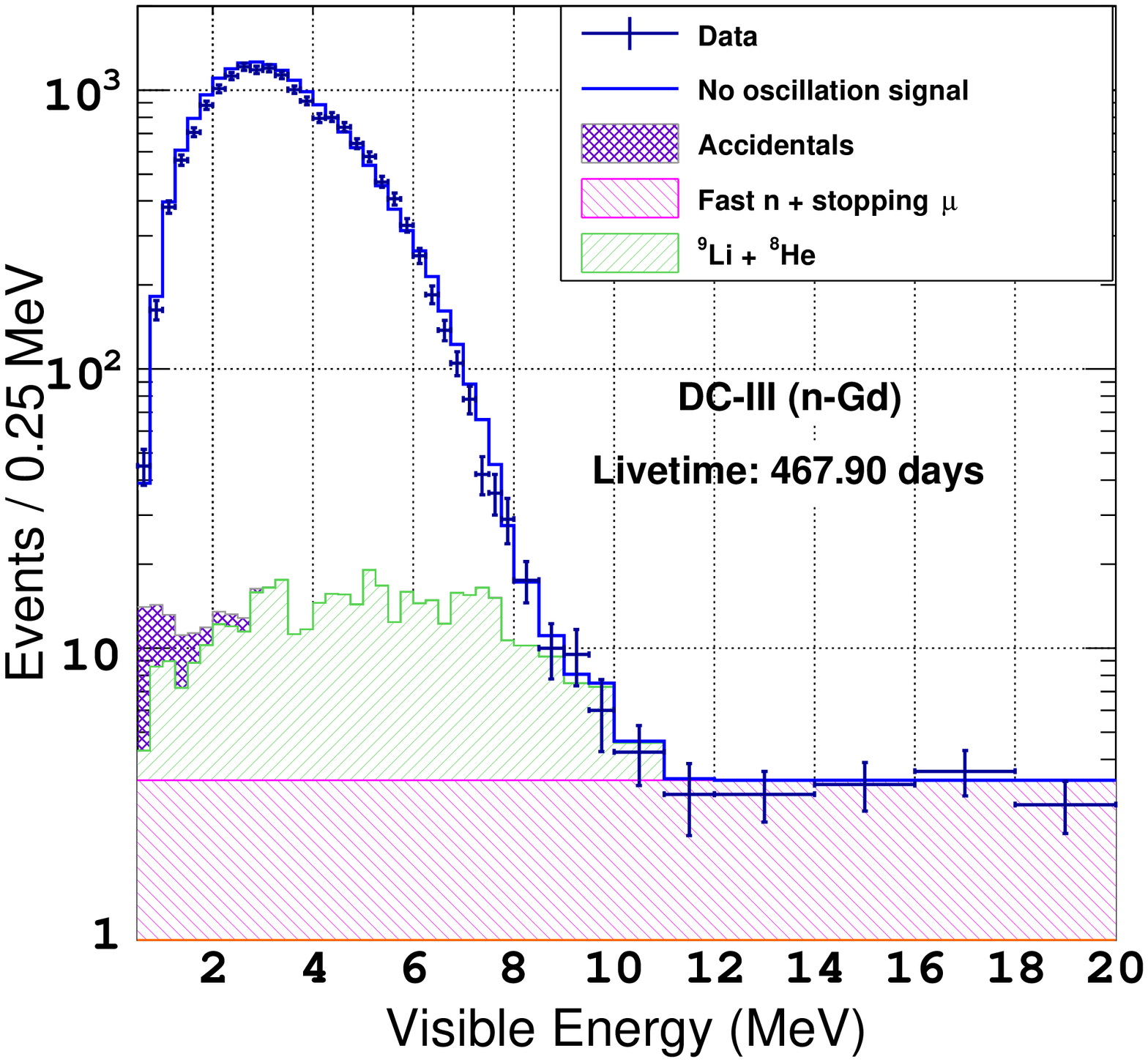}
\includegraphics[width=0.49\textwidth]{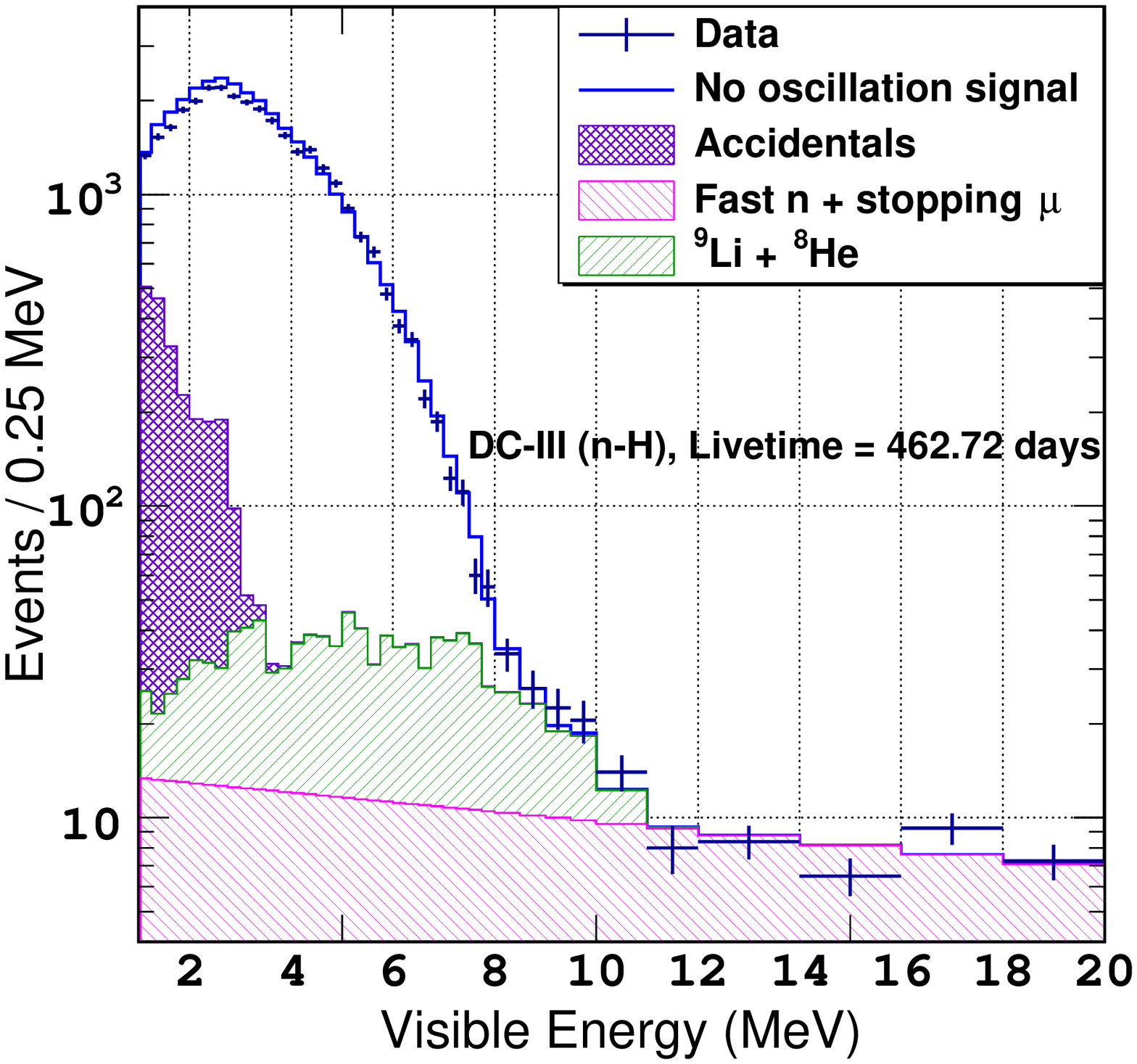}
\end{center}
\caption{\small{The stacked IBD and background spectra for the n-Gd (left) and n-H (right) analysis. The high suppression of accidentals of the n-H analysis can be seen, making the background levels of the two analysis almost comparable.}}
\label{fig:IBDspectra}
\end{figure}

All the signal and background components of the n-Gd and n-H analysis are summarized in Fig.~\ref{fig:IBDspectra}, where the stacked prompt spectra are shown, while table~\ref{tab:IBDandBKG} shows the integrated rate per day. On the last line of this table, the signal over background ratio is presented, where an improvement over each new released can be seen.

\begin{table}[htbp]
\centering
\caption{\small{Comparison between the rates per day [day$^{-1}$] of the IBD prediction, background components and data for the DC data-sets released so far. The last line is the signal over background ratio.}}
\resizebox{\columnwidth}{!}{%
\begin{tabular}{lccccc}
\hline
\hline
						&	Gd-I				&	Gd-II				&	H-II					&	Gd-III						&	H-III					\\
\hline
IBD Prediction			&	44.9 $\pm$ 1.7	&	36.19 $\pm$ 0.96		&	73.68 $\pm$ 2.28		&	38.05 $\pm$ 0.69				&	66.0 $\pm$ 1.3			\\
Accidentals				&	0.33 $\pm$ 0.03	&	0.261 $\pm$ 0.002	&	73.45 $\pm$ 0.16		&	0.070 $\pm$	0.003			&	4.33 $\pm$ 0.01			\\
Cosmogenic				&	2.3 $\pm$ 1.2	&	1.25 $\pm$ 0.54		&	2.8 $\pm$ 1.2		&	0.97$^{ + 0.41}_{ - 0.16}$	&	0.95$^{+0.57}_{-0.33}$	\\
Correlated				&	0.83 $\pm$ 0.38	&	0.67 $\pm$ 0.20		&	3.17 $\pm$ 0.54		&	0.604 $\pm$ 0.051			&	1.55 $\pm$ 0.15			\\
\hline
Data						&	42.6				&	36.2					&	151.12				&	37.66						&	69.9						\\
S/B						&	11.3				&	15.6					&	0.9					&	21.9							&	9.2						\\
%
\hline
\hline
\end{tabular}%
}
\label{tab:IBDandBKG}
\end{table}


\subsection{Oscillation Analysis}
\label{sec:OscAna}



\begin{figure}[htbp]
\begin{center}
\includegraphics[width=0.49\textwidth]{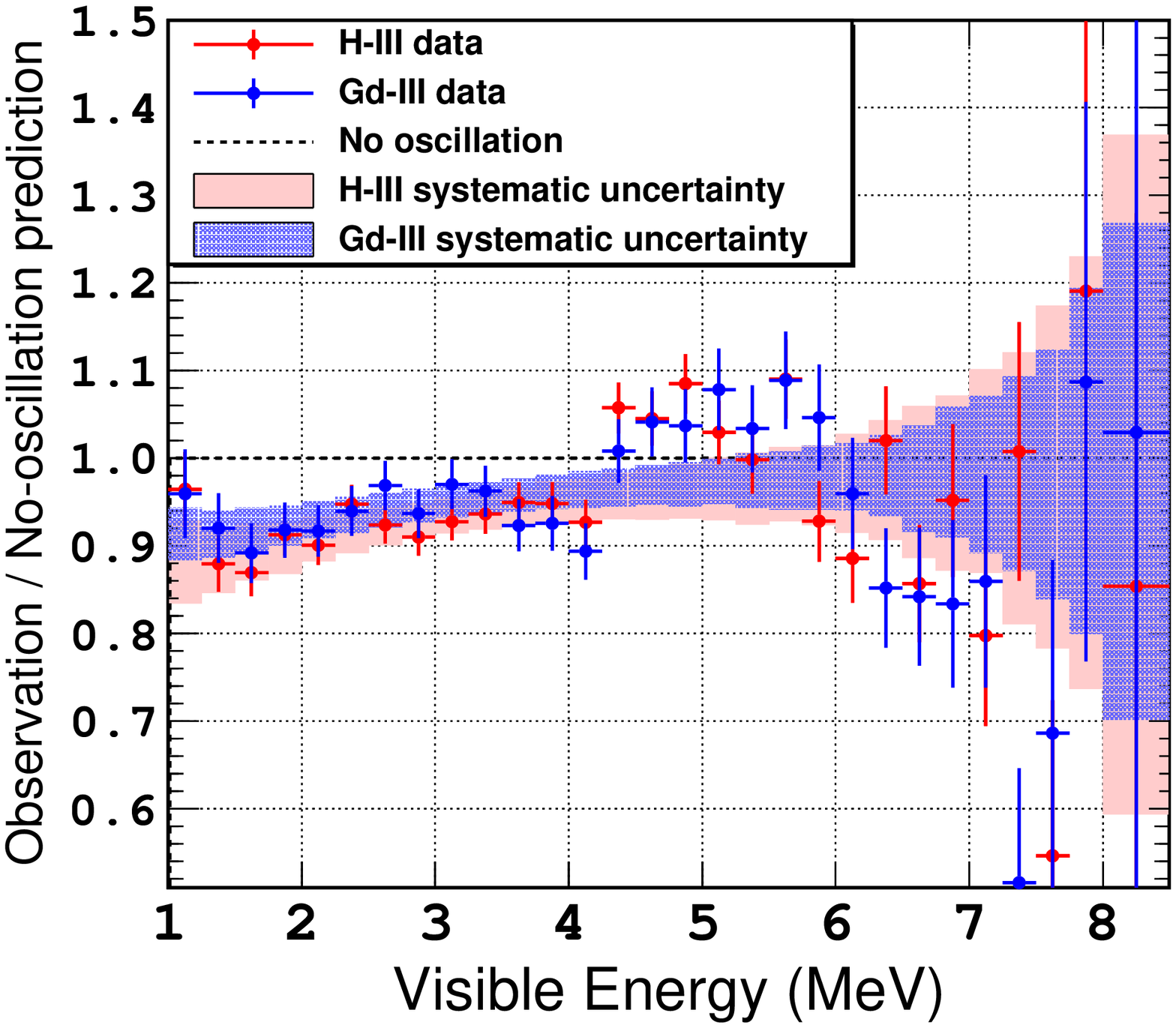}
\includegraphics[width=0.49\textwidth]{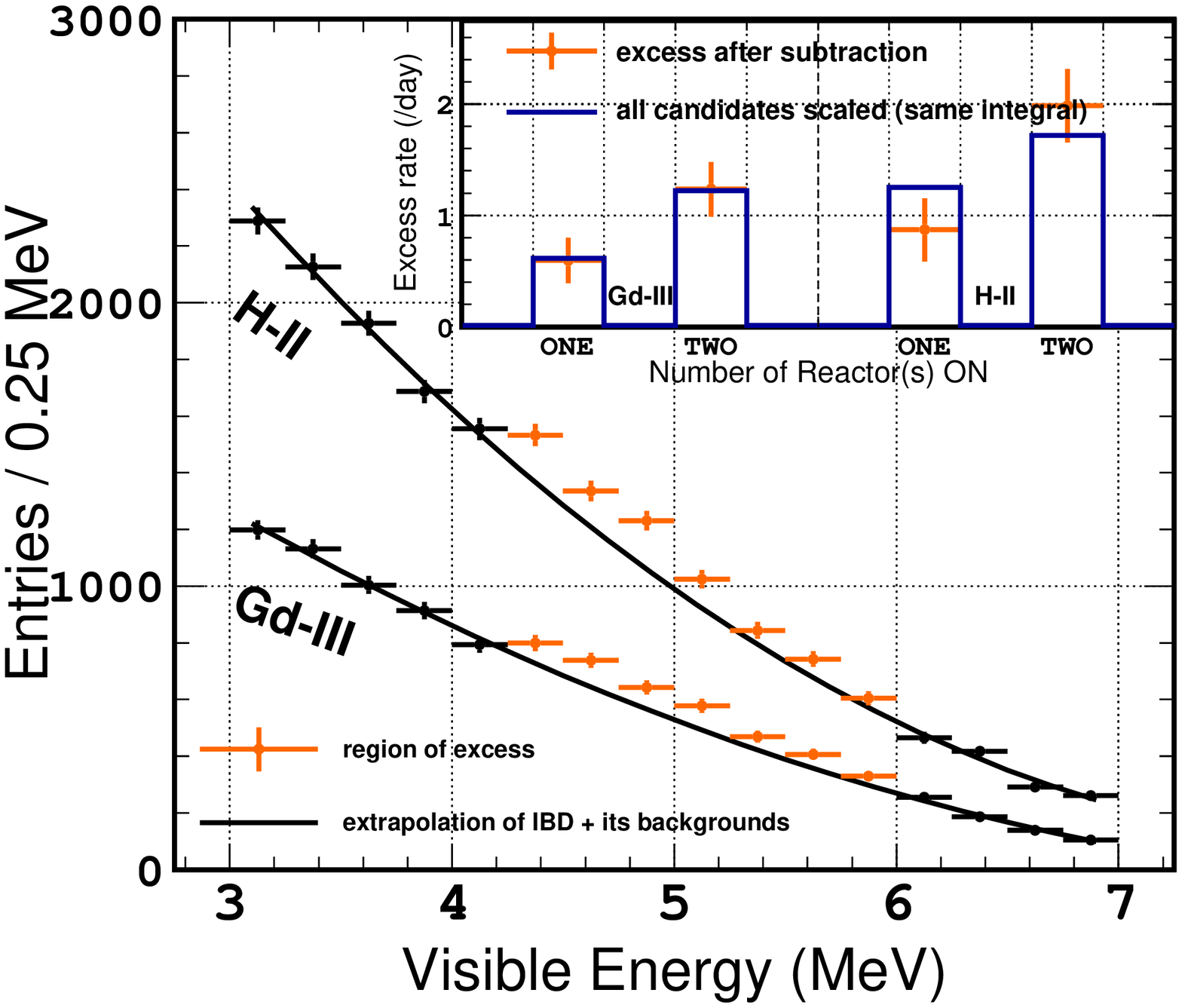}
\end{center}
\caption{\small{The reactor neutrino spectrum ``bump''. Left: Ratio between the best-fit and non-oscillated spectra for both n-Gd and n-H analysis, where a structure (``bump'') can be seen between 4 and 6~MeV. Right: Zoom of the prompt IBD energy spectra for the Gd-III and H-II selection methods, where a exponential fit was done without taking into account the bins between 4 and 6~MeV, showing an excess in this region. The inset shows the correlation of the excess with the reactor power and analysis method (statistics).}}
\label{fig:SpecExcess}
\end{figure}

Since its first publication, DC uses a spectral shape and rate fit of the data, prediction and backgrounds to measure the value of $\sin^22\theta_{13}$. It is a regular $\chi^2$ method where covariances matrices for the systematic shape uncertainties and pull terms for the known rates and variables. On the last n-Gd DC publication (Gd-III), this method resulted in 
\begin{equation}
\sin^22\theta_{13} = 0.090^{+0.032}_{-0.029}.
\end{equation}
Table~\ref{tab:FitUnc} shows the normalization  uncertainties for this measurement, and also for all the other data-set released so far by the DC collaboration.
\begin{table}[htbp]
\centering
\caption{\small{Comparison of the summary of signal and background normalization uncertainties (in percent [\%]) relative to the signal prediction between the DC data-sets released so far.}}
\begin{tabular}{lccccc}
\hline
\hline
						&	Gd-I		&	Gd-II	&	H-II		&	Gd-III		&	H-III		\\
\hline
Reactor Flux				&	1.8		&	1.7		&	1.8		&	1.7			&	1.7			\\
Detection Efficiency		&	2.1		&	1.0		&	1.6		&	0.6			&	1.0			\\
Cosmogenic				&	2.8		&	1.38		&	1.6		&	+1.1 / -0.4	&	+0.86/-0.5	\\
Correlated				&	0.9		&	0.51		&	0.6		&	0.1			&	0.2			\\
Accidental				&	$<0.1$	&	$<0.1$	&	0.2		&	$<0.1$		&	$<0.1$		\\
Statistics				&	1.6		&	1.1		&	1.1		&	0.8			&	0.6			\\
\hline
\textbf{Total}			&	4.3		&	2.7		&	3.1		&	+2.3 / -2.0	&	+2.3 / -2.2	\\
\hline
\hline
\end{tabular}%
\label{tab:FitUnc}
\end{table}
An interesting feature is illustrated on the left plot of Fig.~\ref{fig:SpecExcess}, that shows the ratio between the best fit and the non-oscillated spectra.
A ``bump'' is present on the 4 to 6~MeV and as the figure shows, this feature is also presented in the independent n-H sample. So far, there is no known background model that could accommodate this difference. Comparing data with the MC model of the energy response at this range, also using n-C captures and $^{12}$B decay, did not show any energy reconstruction misbehaviour as well. However, as shown in the right plot of Fig.~\ref{fig:SpecExcess}, a correlation between the excess and the reactor power was observed, giving an evidence that the current models of reactor neutrino flux might be insufficient. For the Gd analysis, such excess deviates from the prediction with a significance of 3$\sigma$, and has a rate of $0.60\pm 0.20$ ($1.23\pm 0.25$) events per day when one (two) reactor is running. With the addition of the H sample, the excess rate goes to $0.87\pm 0.28$ ($1.99\pm 0.33$) events per day. Although similar results were reported by other experiments, this one from DC remains the only one published by the time of writing this paper\footnote{When revising this paper, a confirmation of the excess by the Daya Bay collaboration was published at~\cite{DBflux}}.

\begin{figure}[htbp]
\begin{center}
\includegraphics[width=0.45\textwidth]{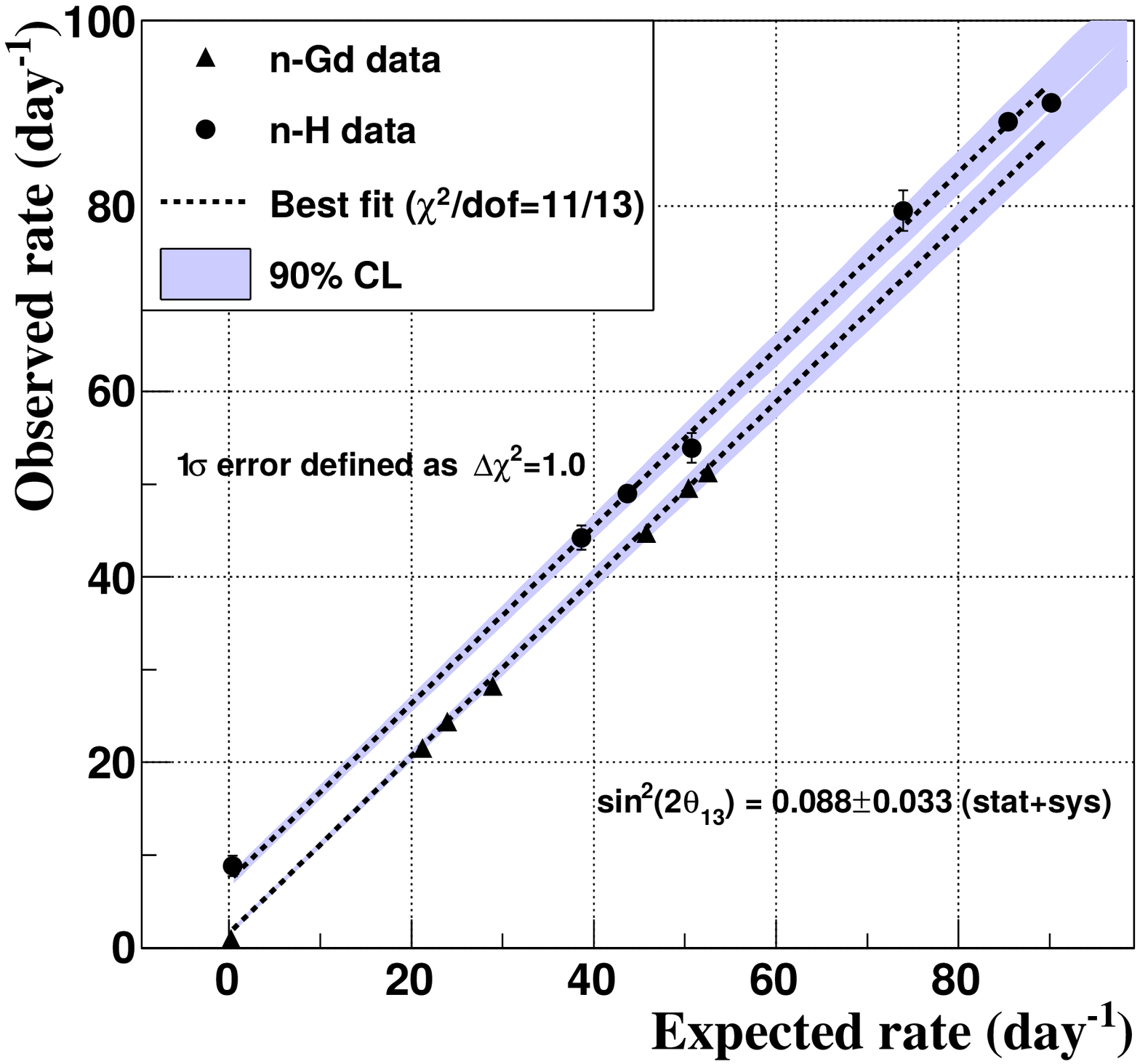}
\includegraphics[width=0.45\textwidth]{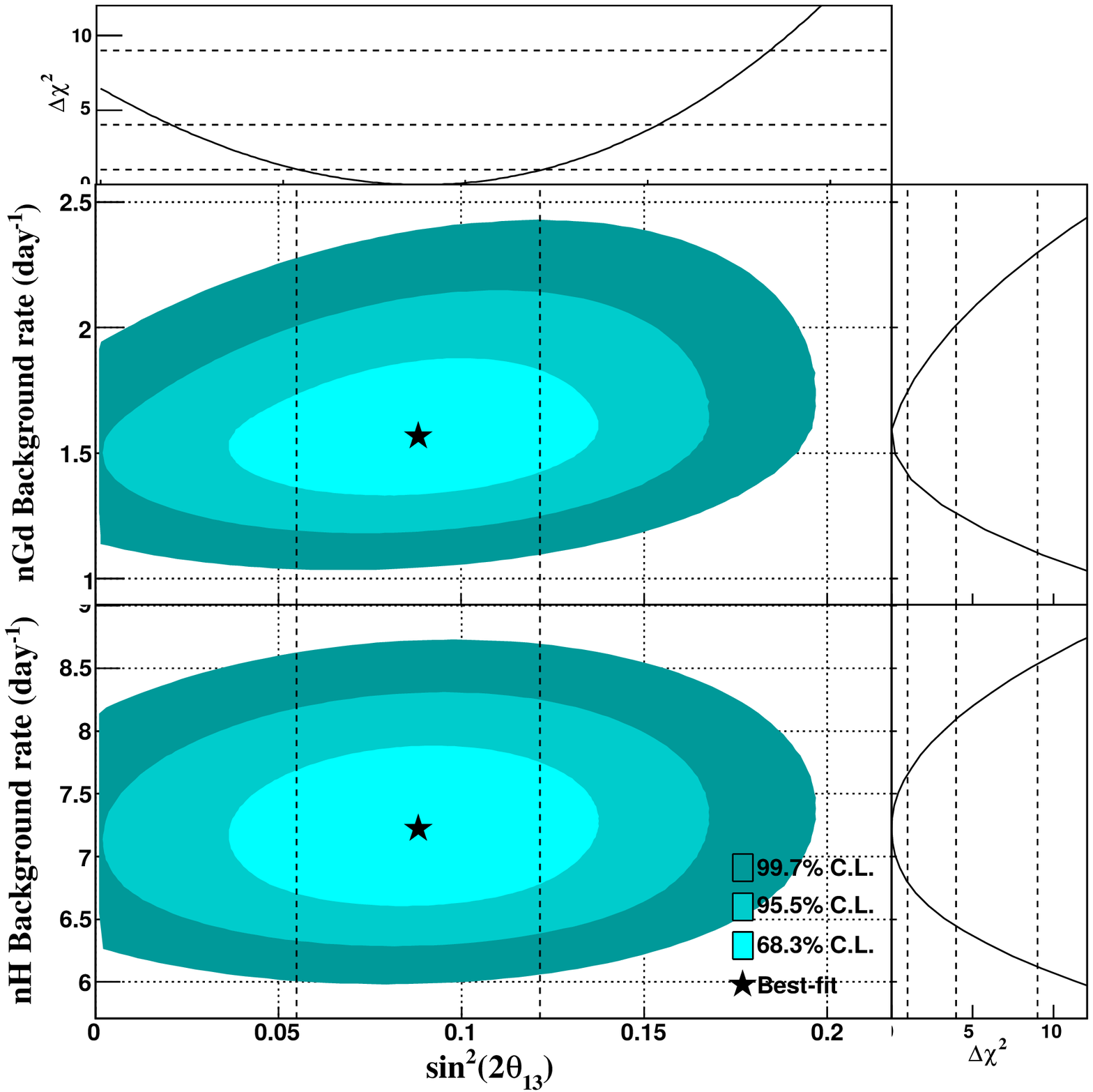}
\end{center}
\caption{\small{Left: Combined n-Gd and n-H RRM fit, where the reactor off data and background expectation are used, assuming uncorrelated background uncertainty and fully correlated flux uncertainty between the data sets. Right: Parameter space of the two fitted parameters, $\sin^22\theta_{13}$ and total background rate. The curves represents the 68.3\%, 95.5\% and 99.7\% C.L., and the one-dimension $\chi^2$ curves are displayed in the upper and lateral insets.}}
\label{fig:RRMfit}
\end{figure}

DC uses another method of oscillation analysis, the so called Reactor Rate Modulation (RRM) fit~\cite{14DCTH13BKI}. 
A fit is performed comparing the predicted IBD rate with the measured one, at different reactor power regimes. It is comparable to a linear-fit, where the mixing angle is given by the slope angle and the total background rate by the intercept, i.e., where none IBD is expected. Therefore, this method is independent of spectral shape information but can still constrain the background rate in the fit. In addition, the RRM fit is robust against the spectrum distortion with a constraint from the Bugey4 measurement.
%
The left plot of Fig.~\ref{fig:RRMfit} shows this method used to fit both the n-Gd and n-H samples simultaneously, which gives 
\begin{equation}
\sin^22\theta_{13} = 0.088 \pm 0.033.
\end{equation}
%
The parameter space of both data-sets can be seen on the right plot of Fig.~\ref{fig:RRMfit}.

\begin{figure}[htbp]
\begin{center}
\includegraphics[width=0.95\textwidth]{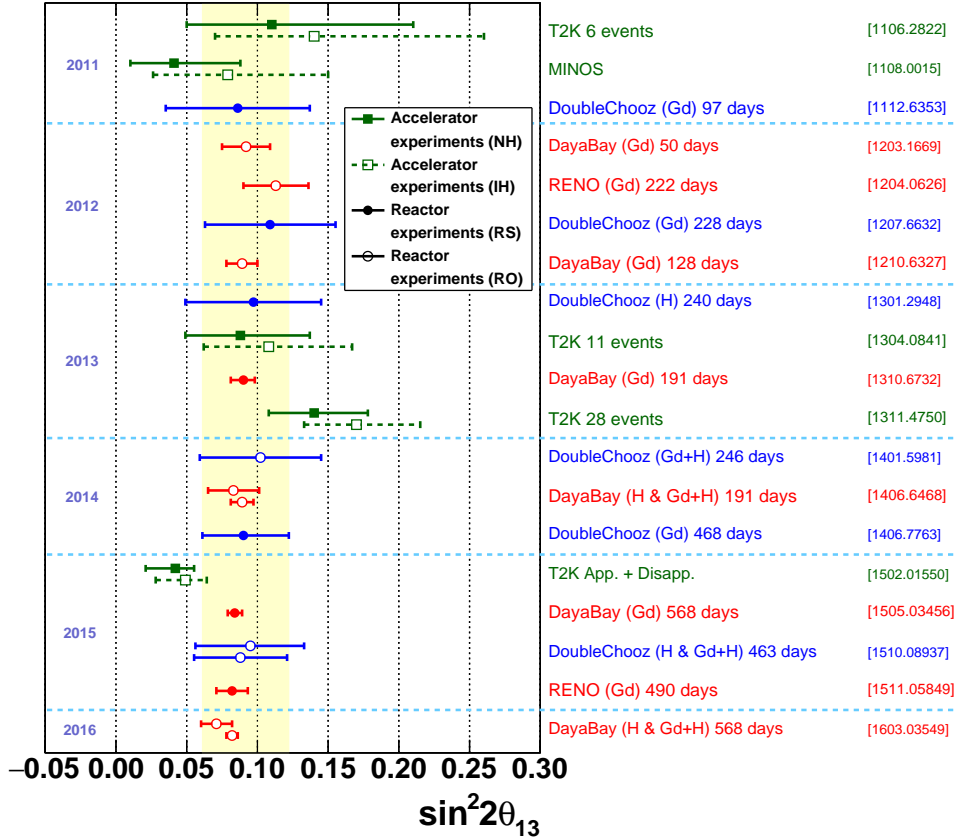}
\end{center}
\caption{\small{Summary of the $\sin^22\theta_{13}$ main measurements released by reactor and accelerator experiments since 2011. The accelerator results are presented in Normal or Inverted Hierarchy (NH and NI respectively), while the reactor results are divided by rate only fit (RO), spectral shape and rate fit (RS) or if the analysis is based on n-Gd, n-H or a combination of both (Gd+H). The number of days represents the live-time of each data-release and when the detectors have different live-time in a same release, the biggest value was taken. The yellow band represents the result with lowest uncertainty delivered by DC so far, the Gd-III analysis. For each value, its arXiv number is given for quick reference.}}
\label{fig:th13summary}
\end{figure}

Figure~\ref{fig:th13summary} is a summary of the main measurement results on $\sin^22\theta_{13}$ released up to now. It shows all the analysis used so far on the reactor experiments (Gd or H analysis with a spectral shape or rate only fit), and the normal or inverted hierarchy (NH and NI respectively) assumptions for the accelerator experiments. 
The figure shows a very good agreement among all experiments and analysis types
\footnote{During the last reviewing process of this paper, the Double Chooz collaboration presented a preliminary near+far result in the Rencontres de Moriond conference, which gives $\sin^22\theta_{13}=0.111\pm0.018$.}. 

%

\begin{figure}[htbp]
\begin{center}
\includegraphics[width=0.6\textwidth]{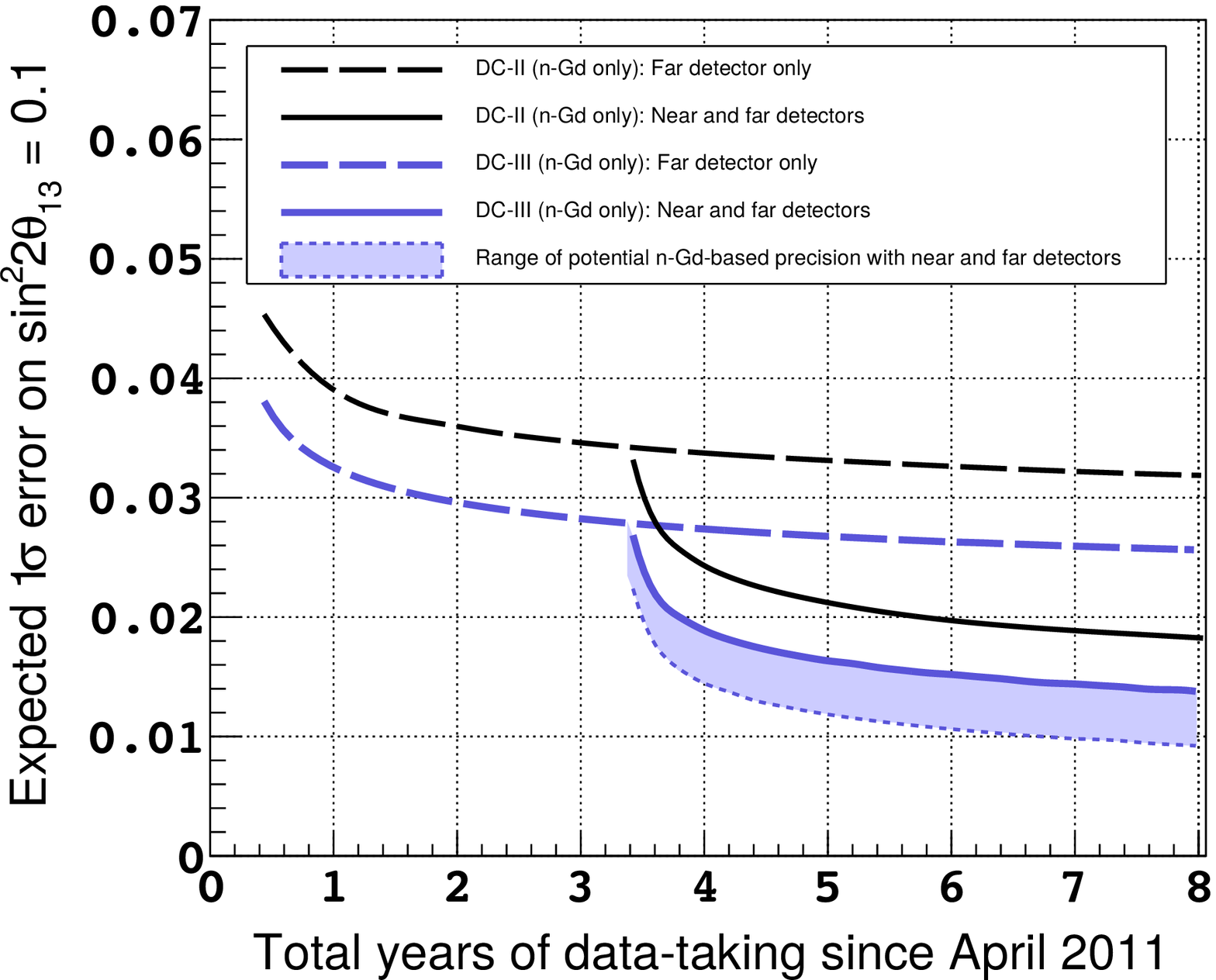}
\end{center}
\caption{\small{Predicted $\sin^22\theta_{13}$ error of the DC experiment, for the n-Gd analysis. The black (light blue) dashed line represents the Gd-II (Gd-III) analysis, while the solid line is the expectation with the ND operation. The shaded region represents the range of improvements expected by the reduction in the systematic uncertainties and the lower edge corresponds to no systematic uncertainty besides the reactor flux.}}
\label{fig:Sensitivity}
\end{figure}

The DC collaboration chose to make a two phase experiment, by first constructing the far detector using the old Chooz experiment laboratory, and after its completion and start of data taking, build the Near Detector (ND) (preceded by tunnel and cave excavation and laboratory built). Figure~\ref{fig:Sensitivity} shows the predicted sensitivity of the experiment, for the n-Gd analysis, with the operation of the ND. In this figure it can also be seen how the improvements on the data analysis from Gd-II (black line) to Gd-III (light blue line) impacted the sensitivity on $\sin^22\theta_{13}$. The shaded region represents the range of improvements expected by the reduction in the systematic uncertainties.
While dominated by statistic uncertainty, the Double Chooz experiment will decrease considerably the systematic uncertainty in the Near+Far phase. In addition, DC is the best candidate to have the lowest reactor flux related uncertainty involved, given that the detectors are placed at nearly equal reactors iso-flux position~\cite{ReactorFluxUncSup}. On Table~\ref{tab:FitUnc}, while the first two lines are expected to vanish almost completely, the background uncertainties, dominant in the new phase, will decrease as well, since its measurement is based on the statistics of the data accumulation.

Finally, besides works on neutrino oscillation, the DC collaboration also released many other interesting physics analysis, such as a test of Lorentz violation~\cite{12DCLo}, ortho-positronium observation~\cite{14DCPs}, and muon capture on light isotopes~\cite{15DCMuCap}.

\section{Summary}
\label{sec:Summary}
The reactor measurement of $\theta_{13}$ is a very important part of the neutrino oscillation studies. 
Double Chooz initiated the field from the era of the Chooz experiment. 
It showed the first indication of the reactor neutrino deficit at a baseline $\sim $1~km and is measuring $\theta_{13}$ using high quality data and analysis techniques. 
In the near future, it will present a precise $\theta_{13}$ using the near+far detector data. 
The high precision $\theta_{13}$ value measured by reactor experiments will become an indispensable piece of information in the next decades when the mass hierarchy and leptonic CP violation are to be measured and insights into the nature are expected to be deepened. 

\section*{Acknowledgements}
The Double Chooz group thanks the French electricity company EDF; 
the European fund FEDER; 
the R\'{e}gion de Champagne Ardenne; the D\'{e}partement des Ardennes; 
and the Communaut\'{e} de Communes Ardenne Rives de Meuse. 
The Double Chooz group acknowledges the support of the CEA, CNRS/IN2P3, 
the computer centre CCIN2P3, and LabEx UnivEarthS in France (ANR-11-IDEX-0005-02); 
the Ministry of Education, Culture, Sports, Science and Technology of Japan (MEXT) and the Japan Society for the Promotion of Science (JSPS); 
the Department of Energy and the National Science Foundation of the United States; 
U.S. Department of Energy Award DE-NA0000979 through the Nuclear Science and Security Consortium; 
the Ministerio de Econom\'{i}a y Competitividad (MINECO) of Spain; the Max Planck Gesellschaft, and the Deutsche Forschungsgemeinschaft DFG, the Transregional Collaborative Research Center TR27, the excellence cluster gOrigin and Structure of the Universeh, and the Maier-Leibnitz-Laboratorium Garching in Germany; the Russian Academy of Science, the Kurchatov Institute and RFBR (the Russian Foundation for Basic Research); the Brazilian Ministry of Science, Technology and Innovation (MCTI), the Financiadora de Estudos e Projetos (FINEP), the Conselho Nacional de Desenvolvimento Cient\'{i}fico e Tecnol\'{o}gico (CNPq), the S\~{a}o Paulo Research Foundation (FAPESP), and the Brazilian Network for High Energy Physics (RENAFAE) in Brazil.

\section*{References}


\begin{thebibliography}{00} 
%
\bibitem{98SK} Super-Kamiokande Collaboration, Phys. Rev. Lett. {\bf 81} (1998) 1562-1567
%
\bibitem{02SNO} SNO Collaboration,   Phys. Rev. Lett. {\bf 89} (2002) 011301
%
%
\bibitem{15Suekane} F.~Suekane, {\it Neutrino Oscillations}:~{\it A practical guide to basics and applications} (Springer), (2015), DOI:~10.1007/978-4-431-55462-2
%
\bibitem{NeMass} Maintz Collaboration, Eur. Phys. J. C {\bf 33} (2004) S805-S807; 
Troitsk Collaboration, Phys. Rev. D {\bf 84} (2011) 112003
%
\bibitem{14Capozzi} F.~Capozzi, G. L.~Fogli, E.~Lisi, A.~Marrone, D.~Montanino, A.~Palazzo, Phys. Rev. D {\bf 89} (2014) 093018 
%
\bibitem{85Jarlskog} C.~Jarlskog, Phys. Rev. Lett. {\bf 55} (1985) 1039 
%
\bibitem{86Fukugita} M.~Fukugita, T.~Yanagida, Phys. Lett. B \textbf{174} (1986) 45 
%
\bibitem{06Losecco} J.~LoSecco, http://www.hep.nd.edu/NuHistpap.pdf
%
\bibitem{ChoozPaloVerde} Chooz Collaboration, Eur. Phys. J. C \textbf{27} (2003) 331-374; 
Palo Verde Collaboration, Phys. Rev. D \textbf{64} (2001) 112001 

\bibitem{99ChoozPaloVerde} Chooz Collaboration, Phys. Lett. B{\bf 466}:415-430,1999; 
Palo Verde Collaboration, Phys.Rev.Lett. {\bf 84} (2000) 3764-3767 
%
\bibitem{98Narayan} M.~Narayan, G.~Rajasekaran, S.~Uma Sanker,  
Phys.Rev. D\textbf{58} (1998) 031301 


\bibitem{01T2KProposal} Y.~Itow, T.~Kajita, K.~Kaneyuki {\it et al.}, hep-ex/0106019 (2001)
%
\bibitem{03KamLAND} KamLAND Collaboration, Phys. Rev. Lett. {\bf 90} (2003) 021802
%
\bibitem{99KR2DET} L. Mikaelyan and V. Sinev, Yad. Fiz., 63, issue 6, p. 1077, 2000. (Physics
of Atomic Nuclei, 63, issue 6, p. 1002, 2000.),  hep-ex/9908047 (1999)
%
\bibitem{03Minakata} H.~Minakata, H.~Sugiyama, O.~Yasuda, K.~Inoue, F.~Suekane, 
Phys. Rev. D \textbf{68} (2003) 033017, Phys. Rev. D \textbf{70} (2004) 059901
%
\bibitem{03Huber} P.~Huber, M.~Lindner, T.~Schwetz, W.~Winter,
Nucl. Phys. B \textbf{665} (2003) 487-519
%
\bibitem{04DCLoI}  Double Chooz Collaboration, hep-ex/0405032 (2004)
%
\bibitem{04White}  K.~Anderson {\it et al.}, hep-ex/0402041 (2004) 
%
\bibitem{06DCLoI} Double Chooz Collaboration,  hep-ex/0606025 (2006)
%
\bibitem{03KR2DET} Yu.~Kozlov, L.~Mikaelyan, V.~Sinev, Yad. Fiz. {\bf 66} (2003) 497
[Phys. At. Nucl. \textbf{66} (2003) 469] 
%
\bibitem{06KASKALoI} KASKA Collaboration, hep-ex/0607013 (2006)
%
\bibitem{DiabloCanyon} http://theta13.lbl.gov/index\_diablocanyon.html
%
\bibitem{03Shaevitz} M.H.~Shaevitz, J.M.~Link, hep-ex/0306031 (2003)
%
\bibitem{07DayaBay} Daya Bay Collaboration, hep-ex/0701029  (2007)
%
\bibitem{10RENO} RENO Collaboration,	arXiv:1003.1391 [hep-ex] (2010)

\bibitem{04USDC} S.~Berridge {et al.}, hep-ex/0410081 (2004)
%
\bibitem{11T2KTh13} T2K Collaboration, Phys. Rev. Lett. \textbf{107} (2011) 041801
%
\bibitem{11MINOSTh13} MINOS Collaboration, Phys. Rev. Lett. \textbf{107} (2011) 181802 
%
\bibitem{12DC1st}  Double Chooz Collaboration,  Phys. Rev. Lett. \textbf{108} (2012) 131801
%
\bibitem{12DayaBay1st}  Daya Bay Collaboration,  Phys. Rev. Lett. {\bf 108} (2012) 171803
%
\bibitem{12RENO1st}  RENO Collaboration,  Phys. Rev. Lett. {\bf 108} (2012) 191802
%
\bibitem{15DayaBay} Daya Bay Collaboration, arXiv:1505.03456 (2015)
%
\bibitem{14T2K} T2K Collaboration,  Phys. Rev. Lett. {\bf 112} (2014) 061802 

\bibitem{15NOVA} NOvA Collaboration, arXiv:1601.05022 [hep-ex] (2016)
%
\bibitem{DCDetector} C.~Anatael for DC Collaboration, Nucl. Instrum. Meth. A \textbf{617} (2010) 473-477; 
F.~Suekane for DC Collaboration, Nucl. Instrum. Meth. A \textbf{623} (2010) 440-441; 
D.~Greiner {\it et al.} Nucl. Instrum. Meth. A \textbf{581} (2007) 139-142; 
A.~Tonazzo {\it et al.} for DC Collbaoration, Nucl. Phys. Proc. Suppl. \textbf{172} (2007) 41-44;
J.V.~Dawson for DC Collaboration, Nuovo Cim. C035N1 (2012) 127-132 
%
\bibitem{DCLS} C.~Buck {\it et al.} Nucl. Phys. Proc. Suppl. \textbf{221} (2011) 372; 
C.~Aberle {\it et al.} JINST \textbf{6} (2011) P11006; 
C.~Aberle {\it et al.} Nucl. Phys. Proc. Suppl. \textbf{448} (2012) 229-232
%
\bibitem{DCPMT} T.~Matsubara {\it et al.}, Nucl. Instrum. Meth. A \textbf{661} (2012) 16-25; 
C.~Bauer {\it et al.}, JINST \textbf{6} (2011) P06008; 
E.~Calvo {\it et al.}, Nucl. Instrum. Meth. A \textbf{621} (2010) 222-230;
J.~Haser {\it et al.}, JINST \textbf{8} (2013) P04029; 
F.~Sato {\it et al.}, Phys. Procedia \textbf{37} (2012) 1164-1170 
%
\bibitem{DCIV} D.~Dietrich {\it et al.}, JINST \textbf{7} (2012) P08012; 
K.~Zbiri, arXiv:1104.4045 [physics.ins-det] (2011)
%
\bibitem{DCReadout} Y.~Abe {\it et al.}, JINST \textbf{8} (2013) P08015; 
F.~Beissel {\it et al.} JINST \textbf{8} (2013) T01003;
T.~Akiri for DC Collaboration, Nucl.Phys.Proc.Suppl. \textbf{458} (2012) 229-232; 
J.~Maeda for DC Collaboration, J.Phys.Conf.Ser. \textbf{331} (2011) 022018; 
T.~Konno {\it et al.}, DOI: 10.1109/NSSMIC.2009.5401999 
%
\bibitem{12DCTH13}  Double Chooz Collaboration, Phys. Rev. D \textbf{86} (2012) 052008  
%
\bibitem{13DCTH13}  Double Chooz Collaboration, Phys. Lett. B \textbf{723} (2013) 66-70
%
\bibitem{14DCTH13IMP}  Double Chooz Collaboration, JHEP \textbf{723} (2014) 086; JHEP \textbf{1502} (2015) 074 
%
\bibitem{15DCTH13} Double Chooz Collaboration, arXiv:1510.08937 [hep-ex] (2015) (accepted for publication on JHEP)
%
\bibitem{DCReactor} C.L.~Jones {\it et al.}, Phys. Rev. D \textbf{86} (2012) 012001
%
\bibitem{NuFlux} K.~Schreckenbach {\it et al.}, Phys. Lett. B \textbf{160} (1985) 325; 
F.~von Feilitzsch and K.~Schreckenbach, Phys. Lett. B \textbf{118} (1982) 162; 
A.~Hahn {\it et al.}, Phys. Lett. B \textbf{218} (1989) 365; 
N.~Haag {\it et al.}, Phys. Rev. Lett. \textbf{112} (2014) 122501
%
\bibitem{Bugey4} Y.~Declais {\it et al.}, Phys. Lett. B \textbf{338} (1994) 383
%
\bibitem{14DCBKG}  Double Chooz Collaboration, Phys. Rev. D \textbf{87} (2013) 011102
%
\bibitem{DCMuonReco}  Double Chooz Collaboration, Nucl. Inst. Meth. Phys. Res. A \textbf{764} (2014) 330-339
%
\bibitem{DBflux}  Daya Bay Collaboration, Phys. Rev. Lett. \textbf{116} (2016) 061801
%
\bibitem{14DCTH13BKI}  Double Chooz Collaboration, Phys. Lett. B \textbf{735} (2014) 51-56 
%
\bibitem{ReactorFluxUncSup} A.~Cucoanes {\it et al.}, arXiv:1501.00356 (2015)
%
\bibitem{12DCLo}  Double Chooz Collaboration, Phys. Rev. D \textbf{86} (2012) 112009
%
\bibitem{14DCPs}  Double Chooz Collaboration, JHEP \textbf{1410} (2014) 32
%
\bibitem{15DCMuCap}  Double Chooz Collaboration, arXiv:1512.07562 (2015)

\end{thebibliography}

\end{document}